\documentclass[%
 reprint,
 amsmath,amssymb,
 aps,
]{revtex4-2}

\usepackage{graphicx}
\usepackage{dcolumn}
\usepackage{subcaption}
\usepackage{bm}

\usepackage{xcolor}
\begin{document}

\preprint{APS/123-QED}

\title{Complete Decomposition of Anomalous Diffusion \\ in Variable speed Generalized L\'evy Walks}

\author{Abhijit Bera}
\affiliation{%
 Department of Physics, University of Houston
}%
\affiliation{
 Texas Center for Superconductivity, University of Houston
}
\author{Kevin E. Bassler}%
\email{bassler@uh.edu}
\affiliation{%
 Department of Physics, University of Houston
}%
\affiliation{
 Texas Center for Superconductivity, University of Houston
}
\affiliation{
 Department of Mathematics, University of Houston
}


\date{\today}

\begin{abstract}
Variable Speed Generalized Lévy Walks (VGLWs) are a class of spatio-temporally coupled stochastic processes that unify a broad range of previously studied models within a single parametrized framework. Their dynamics consist of discrete random steps, or flights, during which the walker’s speed varies deterministically with both the elapsed time and the total duration of the flight. We investigate the anomalous diffusive behavior of VGLWs and analyze it through decomposition into the three fundamental constitutive effects that capture violations of the Central Limit Theorem (CLT):
the Joseph effect, reflecting long-range increment correlations, the Noah effect, arising from heavy-tailed step-size distributions with infinite variance, and the Moses effect, associated with statistical aging and non-stationarity. Our results show that anomalous diffusion in VGLWs is typically generated by a nontrivial combination of all three effects, rather than being attributable to a single mechanism. Strikingly, we find that within the VGLW framework the Noah exponent 
$L$, which quantifies the strength of the Noah effect, is unbounded from above, revealing a richer and more extreme landscape of anomalous diffusion than in previously studied Lévy-walk-type models.

\end{abstract}

\maketitle


\section{Introduction}

Diffusive processes describe the spreading of particles over time as a result of the accumulation of many random steps or increments. If these increments are assumed to be independent, identically distributed, and possess finite variance, then the Central Limit Theorem (CLT)~\cite{laplace1810approximations,fischer2011history} applies. In this case, the distribution of particle displacements converges to a Gaussian distribution as the number of increments increases. Moreover, the variance of the distribution grows linearly with time. A diffusive process that satisfies the CLT is referred to as a normal diffusive process. A classical example is the Wiener process~\cite{brown2022miscellaneous}.

However, numerous experimental systems exhibit anomalous diffusion, in which the mean squared displacement (MSD) of particles scales with time as  
\begin{equation}
\langle x^2(t)\rangle \propto t^{2H}, \quad H \neq \tfrac{1}{2}    
\end{equation}
Various stochastic continuous-time random walks (CTRWs) models have been proposed to capture the essence of such anomalous dynamics~\cite{klafter1987stochastic,akimoto2013distributional,akimoto2014phase,schulz1997anomalous,chen2017anomalous,lim2002self,jeon2014scaled,thiel2014scaled}. In this paper, we focus on \textit{Variable Speed Generalized L\'evy Walks} (VGLWs)~\cite{albers2018exact,albers2022nonergodicity}. VGLWs provide a broad paradigm that encompasses many previously studied CTRW-type models. In the VGLW framework, particles take independent steps of random duration drawn from a power-law distribution. The distance traveled in each step scales as a power of the step duration, while the direction of motion is random. Importantly, during each step the particle’s velocity evolves as a deterministic function of both the step duration and the elapsed time. 


Anomalous diffusion arises when the CLT is violated. It can be decomposed into three fundamental constitutive effects, each associated with a distinct mechanism of CLT violation~\cite{chen2017anomalous}: correlations between increments, known as the \textit{Joseph effect}; increment distributions with infinite variance, termed the \textit{Noah effect}; and non-stationarity or statistical aging of the increments, referred to as the \textit{Moses effect}. These three mechanisms---either independently or in combination---can generate anomalous diffusion. Each effect can be quantified by an independent scaling exponent that characterizes its relative strength. This decomposition provides a powerful framework for identifying the origin of anomalous behavior, even when the details of the underlying microscopic dynamics are inaccessible. In particular, it enables one to isolate the dominant mechanisms in experimental systems and to construct more accurate models of real-world processes.


The complete decomposition of anomalous diffusion was first carried out in~\cite{chen2017anomalous} for a variety of stochastic processes. In~\cite{meyer2018anomalous}, the framework was extended to an aging deterministic system. It was further developed in~\cite{aghion2021moses} for Generalized L\'evy Walks (GLWs) in the big-jump regime, where the distribution of jump times has an infinite mean. In~\cite{vilk2022unravelling}, the decomposition was applied to diverse real-world systems, ranging from the macroscopic motion of molecules to large-scale phenomena such as bird migration. Studies have employed this decomposition to examine different real-world processes, including finance and biological systems~\cite{trillot2025evidence,meyer2023return,barraza2025non,zamani2021anomalous,salek2024statistical,salek2024equity}.  
In addition to analytical and empirical approaches, machine learning techniques have also been proposed to infer diffusion models directly from single-particle trajectories and to estimate all associated scaling exponents~\cite{meyer2022decomposing,munoz2021objective,argun2021classification,garibo2021efficient,malinowski2025cinnamon}.


In this paper, we analytically decompose the anomalous diffusion of VGLWs across their entire three-dimensional parameter space. While a portion of this parameter space has been analyzed previously~\cite{aghion2021moses}, our work provides a comprehensive treatment. We uncover a variety of dynamical phases, each characterized by a distinct combination of the three fundamental constitutive effects that collectively generate the anomalous diffusive behavior. A key finding is that within the VGLW framework there exists \textit{no upper bound} to the magnitude of the Noah effect. In earlier studies, the maximum value of the Noah exponent $L$, which quantifies the Noah effect, was found to be less than or equal to unity. By contrast, our analysis shows that in VGLWs the exponent $L$ can grow without bound, revealing a richer and more extreme landscape of anomalous diffusion than previously recognized.

\section{Variable speed Generalized L\'evy Walks (VGLWs)}

Anomalous diffusion is observed across a wide range of real-world systems 
\cite{hofling2013anomalous,metzler2014anomalous,metzler2019brownian,oliveira2019anomalous,
sabri2020elucidating}. Examples include intraday price fluctuations in financial markets 
\cite{chen2017anomalous,bassler2007nonstationary,seemann2012ensemble}, particle motion in 
crowded intracellular environments \cite{sabri2020elucidating}, cold atoms in dissipative 
optical lattices \cite{dechant2012anomalous}, and blinking quantum dots 
\cite{plakhotnik2010anomalous,margolin2004aging}.

A wide variety of stochastic models have been developed to explain such behavior. 
CTRWs \cite{montroll1965random} and their 
extensions are among the most widely used. In CTRWs, Gaussian spatial increments occur at 
random times drawn from a heavy-tailed (power-law decaying) waiting-time distribution . L\'evy flights extend 
this framework by allowing step lengths to follow a heavy-tailed distribution 
\cite{shlesinger1982random,davey1991mandelbrot}. L\'evy walks constitute a further 
generalization, introducing a deterministic coupling between the displacement and the 
duration of each step \cite{shlesinger1993strange,klafter1996beyond,zaburdaev2015levy}. 
In standard L\'evy walks, particles move at constant velocity during each step, and the 
step durations are heavy-tailed 
\cite{zumofen1993levy,zumofen1993scale}. In GLWs
\cite{shlesinger1987levy}, the velocity is nonlinearly coupled to the step duration, 
though it remains constant throughout each step.

VGLWs \cite{albers2018exact,bothe2019mean} are a more general space--time coupled CTRWs. The 
process consists of independent steps whose durations follow a heavy-tailed distribution 
that decays as $\tau^{-\gamma-1}$. In this work, we take the step-time distribution to be
\begin{equation}
\label{original_time_distribution}
\psi(\tau) = \gamma \tau_0^\gamma \tau^{-1-\gamma} 
\, \Theta(\tau \ge \tau_0),
\end{equation}
where $\tau_0$ is the minimal step duration. During each step, the walker moves in a 
random direction with a deterministic time-dependent speed,
\begin{equation}
\label{Variable_speed}
\mathrm{v}_{\nu,\eta}(\tau,t') 
= \eta c \, \tau^{\nu-\eta} t'^{\,\eta-1},
\end{equation}
where $\nu$, $\eta$, $\gamma$ are positive definite parameters that affect scaling properties and $c$ is a positive definite constant that does not affect the scaling properties. By tuning $\eta$, the 
VGLW model interpolates between several well-known processes:  
$\eta = 1$ recovers GLWs \cite{albers2018exact,albers2022nonergodicity};  
$\eta = \nu$ yields the Drude model \cite{schulz1997anomalous,benkadda1998chaos};  
and the limits $\eta \to 0$ and $\eta \to \infty$ correspond to jump--wait--jump and 
wait--jump--wait random walks, respectively.

The MSD of VGLWs was computed in \cite{bothe2019mean}, and the displacement propagator 
for the regime $\gamma < 1$, where dynamics are dominated by the longest flight, was 
derived in \cite{vezzani2020rare}. Owing to their intrinsic space--time coupling and 
flexible parametrization, VGLWs provide a unified and versatile framework for modeling 
anomalous transport across physical, biological, and financial systems.

\section{Constitutive Exponents of Anomalous Diffusion in VGLWs}{\label{3rd_section}}

Diffusion in VGLWs is characterized by the mean squared displacement (MSD),
\begin{equation}
\label{MSD}
\langle x^2(t) \rangle \equiv \langle [x(t) - x(0)]^2 \rangle,
\end{equation}
whose scaling behavior
\begin{equation}
\langle x^2(t) \rangle \sim t^{2H},
\end{equation}
defines the \textit{Hurst exponent} $H$~\cite{alexander1969comments,mandelbrot2002gaussian}.  
For processes obeying the Central Limit Theorem (CLT), $H = \tfrac{1}{2}$ (normal diffusion).  
When $H \neq \tfrac{1}{2}$, the CLT is violated and the process exhibits anomalous diffusion.

As discussed in the Introduction, violations of the CLT arise from three constitutive mechanisms:
(i) long-range correlations (Joseph effect),
(ii) heavy-tailed increment distributions (Noah effect), and
(iii) temporal non-stationarity or aging (Moses effect).
We now formalize these effects for VGLWs by defining the exponents $J$, $L$, and $M$ and the statistical quantities used to extract them.

\subsection{Joseph Exponent $J$}
\label{subsec:joseph}

The Joseph effect quantifies long-range correlations between increments.  
It can be calculated from the scaling behaviour of time-averaged mean squared displacement (TAMSD),
\begin{equation}
\label{TAMSD_J}
\Big\langle \overline{x^2(t,\Delta)} \Big\rangle
=
\left\langle 
\frac{1}{t-\Delta} \int_{0}^{t-\Delta}
\big[x(t_0 + \Delta) - x(t_0)\big]^2 \, dt_0
\right\rangle ,
\end{equation}
which scales as
\begin{equation}
\label{J_scaling}
\Big\langle \overline{x^2(t,\Delta)} \Big\rangle
\sim t^{\,2L + 2M - 2} \, \Delta^{2J}.
\end{equation}
The exponent satisfies $0 \le J \le 1$, with  
$J=0$ for fully anti-correlated increments and $J=1$ for fully correlated ones.  
Alternative estimations of $J$ can be found with rescaled-range statistics (R/S)~\cite{hurst1951long}, wavelet decomposition~\cite{abry1998wavelet}, and DFA~\cite{peng1994mosaic}.

\subsection{Moses Exponent $M$}
\label{subsec:moses}

The Moses effect captures temporal non-stationarity or statistical aging.  
It is obtained from the first moment of the absolute velocity~\cite{aghion2021moses}:
\begin{equation}
\label{Moses_equation}
\langle |v| \rangle \sim t^{\,M - \tfrac{1}{2}}.
\end{equation}
Unlike $J$, the Moses exponent has no intrinsic upper or lower bound.  

\subsection{Noah Exponent $L$}
\label{subsec:noah}

The Noah effect quantifies heavy-tailed increment statistics.  
It is obtained from the second moment of the velocity~\cite{aghion2021moses}:
\begin{equation}
\label{Noah_equation}
\langle v^2 \rangle \sim t^{\,2L + 2M - 2}.
\end{equation}
The Noah exponent satisfies $L \ge \tfrac12$, with larger values corresponding to increasingly heavy-tailed, fluctuation-dominated dynamics.

\subsection{Scaling Relation and Its Application to VGLWs}
\label{subsec:scaling_relation}

The complete decomposition of anomalous diffusion—linking $(J, L, M)$ to the overall diffusion exponent $H$—was introduced in~\cite{chen2017anomalous}.  
Using the scaling Green--Kubo relation~\cite{meyer2017scale}, \cite{aghion2021moses} showed that Generalized L\'evy Walks satisfy
\begin{equation}
\label{scaling_relation}
H = J + L + M - 1.
\end{equation}
The same derivation extends naturally to the VGLW framework, since the scaling form of the Green–Kubo relation remains valid in this process as well.  Thus, once the exponents $J$, $L$, and $M$ are determined from the velocity and TAMSD statistics, the Hurst exponent $H$ follows immediately via Eq.~\eqref{scaling_relation}.

In this work, we compute $L$ and $M$ from the velocity distribution of VGLWs, and determine $J$ from the TAMSD using the scaling Green--Kubo approach~\cite{meyer2017scale,PhysRevX.4.011022}.  
The Hurst exponent is taken from the analytical MSD scaling derived in~\cite{bothe2019mean}.  
A key result of our analysis is that, unlike in previously studied L\'evy-walk-type models, the Noah exponent $L$ is \emph{unbounded}.  
This prediction is confirmed numerically (see sec.~\ref{L_greater_than_1}).

\section{$L$ and $M$ for VGLWs}

According to Eqs.~\ref{Moses_equation} and~\ref{Noah_equation}, determining the exponents
$L$ and $M$ requires evaluating the first and second moments of the absolute velocity.
In principle, these moments can be obtained by integrating the velocity propagator
$p(\mathrm{v},t)$. However, the explicit form of this propagator depends on the parameter regime,
making a direct calculation cumbersome. A detailed discussion of $p(\mathrm{v},t)$ is provided
in Appendix~\ref{Velocity_propagator_vglw_definition}.

A more compact and unified approach is to use the joint distribution of the total step
duration $\tau$ and the elapsed time within the current step $t'$ at a fixed observation
time $t$, denoted by $p(t,\tau,t')$. This representation allows all parameter regimes to be
treated within a single analytical framework.

The distribution $p(t,\tau,t')$ can be expressed as a sum over all possible step numbers,
\begin{equation}
\label{p_sum}
p(t,\tau,t') = \sum_{n=1}^\infty p_n(t,\tau,t'),
\end{equation}
where $p_n(t,\tau,t')$ is the probability that the process is in its $n$-th step at time $t$,
the intended duration of that step is $\tau$, and the elapsed time within that step is $t'$.
Conditioning on being in the $n$-th step gives
\begin{equation}
\begin{aligned}
p_n(t,\tau,t')
&= \Big\langle \delta(\tau - t_n)\,
      \delta\!\left[t' - \Big(t - \sum_{i=1}^{n-1} t_i\Big)\right] \\
&\qquad\qquad \times
      I\!\left(\sum_{i=1}^{n-1} t_i < t < \sum_{i=1}^n t_i\right)
      \Big\rangle ,
\end{aligned}
\end{equation}
where $t_i$ is the duration of the $i$-th step, and $I(\cdot)$ is the indicator function whose value is 1 if the argument is true otherwise 0.

Taking Laplace transforms with respect to the variables
$t\to z$, $\tau\to s$, and $t'\to u$, one obtains the closed-form expression
\cite{godreche2001statistics,albers2022nonergodicity}
\begin{equation}
\label{laplace_p(t,tau,t_prime)}
\widetilde{p}(z,s,u) 
= \sum_{n=1}^\infty \widetilde{p}_n(z,s,u)
= \frac{1}{u+z}\,
  \frac{\widetilde{\psi}(s) - \widetilde{\psi}(s+u+z)}
       {1 - \widetilde{\psi}(z)}.
\end{equation}

Performing the inverse Laplace transforms step by step
\cite{albers2022nonergodicity}, we obtain
\begin{equation}
\widetilde{p}(z,\tau,u)
= \psi(\tau)\,
  \frac{1 - e^{-(u+z)\tau}}{(u+z)\,[1-\widetilde{\psi}(z)]},
\end{equation}
and subsequently
\begin{equation}
\widetilde{p}(z,\tau,t')
= \psi(\tau)\,
  \frac{e^{-z t'}}{1-\widetilde{\psi}(z)}\,
  \Theta(\tau - t'),
\end{equation}
so that the full distribution is
\begin{equation}
\label{p_tau_t_prime}
\begin{aligned}
p(t,\tau,t')
  &= \psi(\tau)\;
     \mathcal{L}^{-1}\!\left[\frac{1}{1 - \widetilde{\psi}(z')}\right]
     \Theta(t - t')\, \Theta(\tau - t'),
\end{aligned}
\end{equation}
where $z'$ is the Laplace variable corresponding to $t - t'$.

The inverse Laplace term depends on whether the mean step duration is finite or infinite:

\paragraph{Case $\gamma < 1$: divergent mean-step duration}
\begin{equation}
\label{R(t)_g_l_1}
\mathcal{L}^{-1}\!\left[\frac{1}{1 - \widetilde{\psi}(z')}\right]
\approx \frac{(t - t')^{\gamma - 1}}
            {|\Gamma(1-\gamma)|\, \Gamma(\gamma)\, t_0^\gamma}.
\end{equation}

\paragraph{Case $\gamma > 1$: finite mean-step duration}
\begin{equation}
\label{R(t)_g_g_1}
\mathcal{L}^{-1}\!\left[\frac{1}{1 - \widetilde{\psi}(z')}\right]
= \frac{\gamma - 1}{\gamma\, t_0}.
\end{equation}

Using $p(t,\tau,t')$, the velocity moments follow as
\begin{equation}
\label{first_moment}
\begin{aligned}
\langle |\mathrm{v}| \rangle
  &= \int\!\!\int |\mathrm{v}|\, p(t,\tau,t') \, d\tau\, dt' \\
  &= \int\!\!\int |\eta c\, \tau^{\nu-\eta} t'^{\,\eta-1}|\;
     p(t,\tau,t')\, d\tau\, dt' ,
\end{aligned}
\end{equation}
\begin{equation}
\label{second_moment}
\begin{aligned}
\langle \mathrm{v}^2 \rangle
  &= \int\!\!\int \mathrm{v}^2\, p(t,\tau,t')\, d\tau\, dt' \\
  &= \int\!\!\int \eta^2 c^2\, \tau^{2\nu-2\eta} t'^{\,2\eta-2}\;
     p(t,\tau,t')\, d\tau\, dt'.
\end{aligned}
\end{equation}

In the following sections, we evaluate these expressions separately for the cases
$\gamma < 1$ and $\gamma > 1$, and derive the corresponding Moses and Noah
exponents $M$ and $L$.

\subsection{$\gamma<1$}
In this case, we can find the first moment of absolute velocity, using Eq.~\ref{first_moment}, as
\begin{eqnarray}
 \langle |\mathrm{v}| \rangle&=&\int_{0}^t \int_{t^\prime}^{\infty}\frac{\gamma}{|\Gamma(1-\gamma)|\Gamma(\gamma)}\eta c \tau^{\nu-\eta}t^{\prime\eta-1}\tau^{-1-\gamma}\nonumber \\
 &&\times (t-t^\prime)^{\gamma-1}\Theta\left(\tau-\tau_0\right) d \tau d t^\prime
 \end{eqnarray}
The above integration with respect $\tau$ is only valid when $\nu<\gamma+\eta$. Moreover, we can expand $(t-t^\prime)^{\gamma-1}$ as $t^{\gamma-1}\left(1-\frac{(\gamma-1)t^{\prime}}{t}+\frac{(\gamma-1)(\gamma-2)t^{\prime2}}{2t^2}+O\left(\left(\frac{t^\prime}{t}\right)^2\right)\right)$.

The upper limit of this integral would yield $t^{\nu-1}$ from all terms of the expansion. For the lower limit, we are only keeping the larger term, $t^{\gamma-1}$.
\begin{equation}
\langle|\mathrm{v}|\rangle\approx\frac{\gamma\eta c}{|\Gamma(1-\gamma)|\Gamma(\gamma)(\eta+\gamma-\nu)}(kt^{\nu-1}-\frac{t^{\gamma-1}\tau_0^{\nu-\gamma}}{\nu-\gamma})     
\end{equation}
where $k$ is a constant and $k\approx\frac{1}{\nu-\gamma}-\frac{\gamma-1}{\nu-\gamma+1}+\frac{(\gamma-1)(\gamma-2)}{2(\nu-\gamma+2)}$.

We can also generate the second moment of $\mathrm{v}$ from Eq.~\ref{second_moment}, as
\begin{eqnarray}
\langle \mathrm{v}^2 \rangle&=&\int_{0}^t \int_{t^\prime}^{\infty}\frac{\gamma\eta^2 c^2}{|\Gamma(1-\gamma)|\Gamma(\gamma)} \tau^{2\nu-2\eta} t^{\prime2\eta-2}\tau^{-1-\gamma}\nonumber\\
&&\times(t-t^\prime)^{\gamma-1}\Theta\left(\tau-\tau_0\right) d \tau d t^\prime
\end{eqnarray}
The above integration with respect $\tau$ is only valid when $2\nu<\gamma+2\eta$. We can also expand $(t-t^\prime)^{\gamma-1}$ for this integral also.

The upper limit of this integral would yield $t^{2\nu-2}$ from all terms of the expansion. For the lower limit, we are only keeping the larger term, $t^{\gamma-1}$.
\begin{equation}
\langle\mathrm{v^2}\rangle\approx\frac{\gamma\eta^2 c^2}{|\Gamma(1-\gamma)|\Gamma(\gamma)(2\eta+\gamma-2\nu)}(k^\prime t^{2\nu-2}-\frac{t^{\gamma-1}\tau_0^{2\nu-\gamma-1}}{2\nu-\gamma-1})    
\end{equation}
where $k^\prime$ is a constant and $k\approx\frac{1}{2\nu-\gamma-1}-\frac{\gamma-1}{2\nu-\gamma}+\frac{(\gamma-1)(\gamma-2)}{2(2\nu-\gamma+1)}$.

Based on the calculation above, three distinct cases emerge in the regime $\gamma < 1$. The derivation of $L$ and $M$ exponents for this regime was calculated~\cite{aghion2021moses}.

\begin{enumerate}
    \item [A.1.] \textbf{$\gamma > \nu$}: In this case, the moments scale as $\langle|\mathrm{v}|\rangle \propto t^{\gamma - 1}$ and $\langle \mathrm{v}^2 \rangle \propto t^{\gamma - 1}$, leading to $M = \gamma - \frac{1}{2}$ and $L = 1 - \frac{\gamma}{2}$.
    
    \item [A.2.] \textbf{$\gamma < \nu < \frac{\gamma}{2} + \frac{1}{2}$}: Here, we find $\langle|\mathrm{v}|\rangle \propto t^{\nu - 1}$ and $\langle \mathrm{v}^2 \rangle \propto t^{\gamma - 1}$, which results in $M = \nu - \frac{1}{2}$ and $L = \frac{\nu}{2} - \gamma + 1$.
    
    \item [A.3.] \textbf{$\frac{\gamma}{2} + \frac{1}{2} < \nu < \frac{\gamma}{2} + \eta$}: In this region, the scaling becomes $\langle|\mathrm{v}|\rangle \propto t^{\nu - 1}$ and $\langle \mathrm{v}^2 \rangle \propto t^{2\nu - 2}$, yielding $M = \nu - \frac{1}{2}$ and $L = \frac{1}{2}$.
\end{enumerate}

\subsection{$\gamma>1$}
In this subsection, we calculate the $L$ and $M$ exponents for $\gamma>1$ region. Starting from Eq.~\ref{first_moment} and Eq.~\ref{second_moment}, we get
\begin{eqnarray}
 \langle |\mathrm{v}| \rangle&=&\int_{0}^t \int_{t^\prime}^{\infty}(\gamma-1)\eta c \tau^{\nu-\eta} t^{\prime\eta-1}\tau^{-1-\gamma}(t-t^\prime)^{0}\nonumber\\
 &&\times\Theta\left(\tau-\tau_0\right) d \tau d t^\prime
\end{eqnarray} 
The above integration with respect $\tau$ is only valid when $\nu<\gamma+\eta$.
\begin{equation}
\langle |\mathrm{v}| \rangle  \approx\frac{(\gamma-1)\eta c}{(\eta+\gamma-\nu)(\nu-\gamma)}(t^{\nu-\gamma}-t^{0}\tau_0)  
\end{equation}

Similarly,
\begin{equation}
 \langle \mathrm{v}^2 \rangle=\int_{0}^t \int_{t^\prime}^{\infty}(\gamma-1)\eta^2 c^2 \tau^{2\nu-2\eta} t^{\prime2\eta-2}\tau^{-1-\gamma}(t-t^\prime)^{0}   
\end{equation}
The above integration with respect $\tau$ is only valid when $2\nu<\gamma+2\eta$. 
\begin{equation}
\langle \mathrm{v}^2 \rangle\approx\frac{(\gamma-1)\eta^2 c^2}{(2\eta+\gamma-2\nu)(2\nu-\gamma-1)}(t^{2\nu-\gamma-1}-t^{0}\tau_0^{2\nu-\gamma-1})    
\end{equation}

\begin{figure}
	\centering
	\includegraphics[width=0.65\columnwidth]{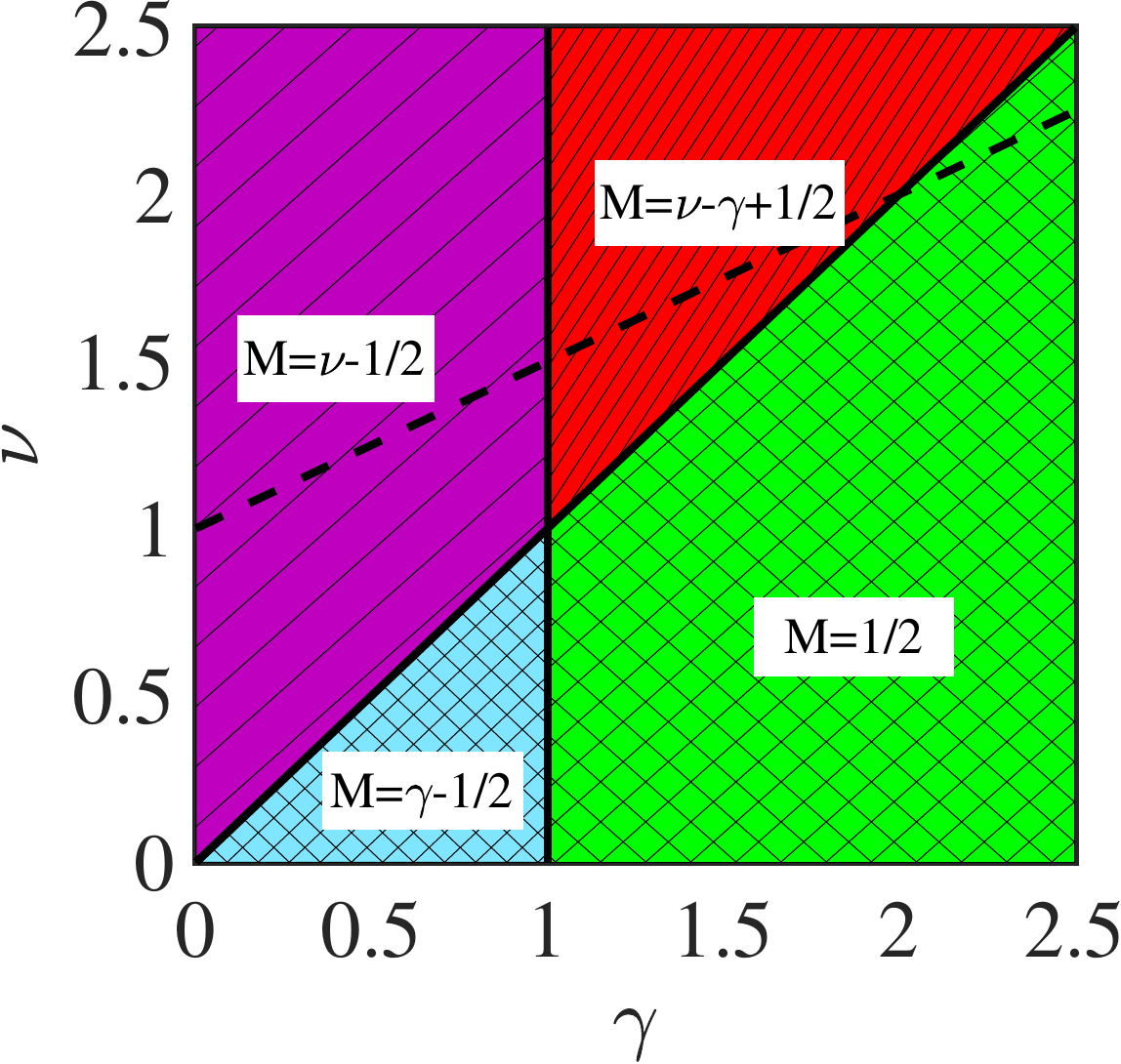}
	\caption{Phase diagram of the Mosses effect for variable speed generalized L\'evy walk. The dotted line marks the onset for the infinite regime for $\eta=1$}
	{\label{Moses_exponent_variable}}
\end{figure}
\begin{figure}
	\centering
	\includegraphics[width=0.65\columnwidth]{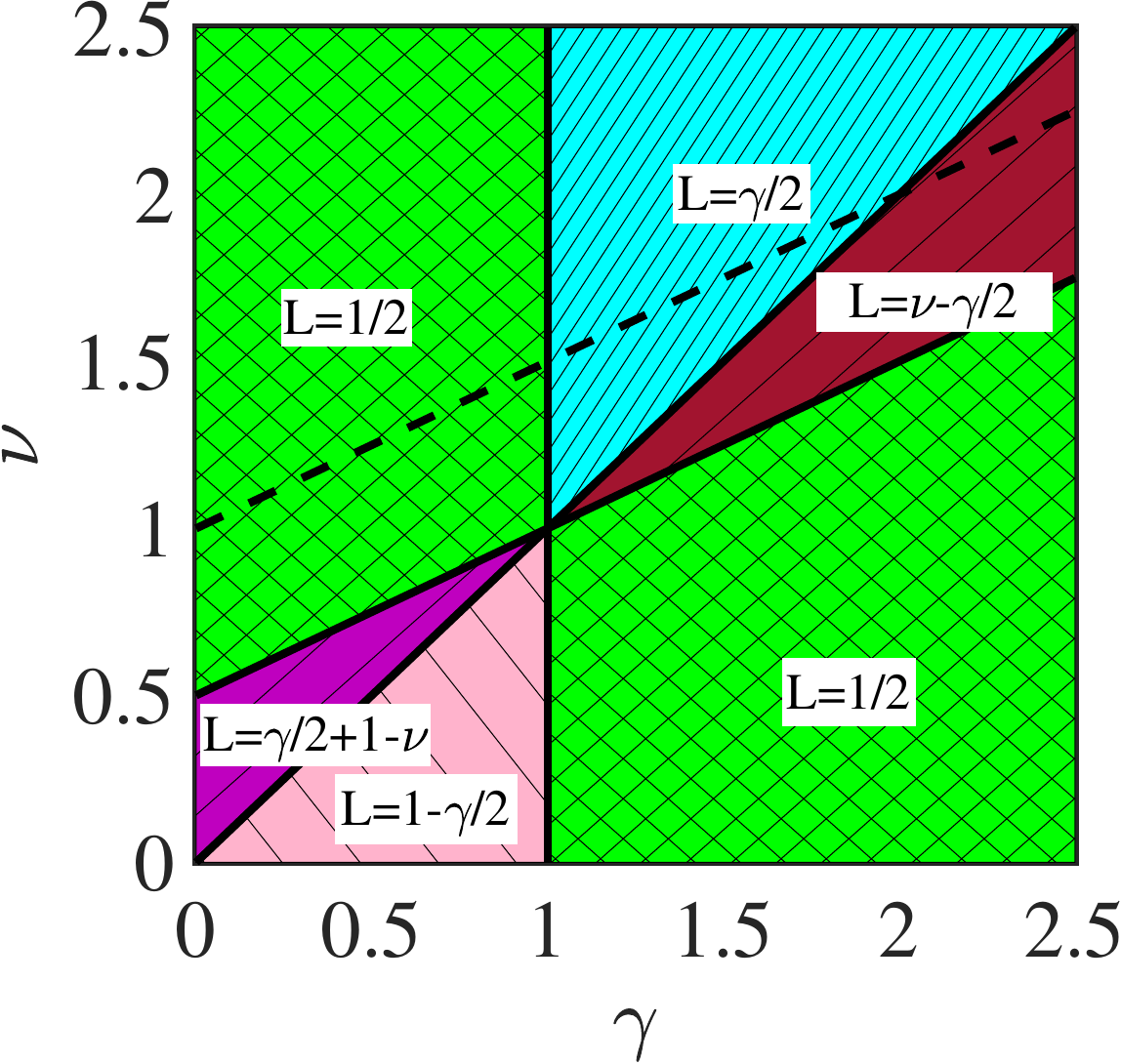}
	\caption{Phase diagram of the Noah effect for variable speed generalized L\'evy walk .The dotted line marks the onset for infinite regime for $\eta=1$}
{\label{Noah_exponent_variable}}
\end{figure}

Here, the behavior of the velocity moments differs significantly from the case of $\gamma<1$. The first moment $\langle|\mathrm{v}|\rangle$ is dominated by the term $t^{\nu - \gamma}$ for $\nu > \gamma$, while the second moment $\langle \mathrm{v}^2 \rangle$ is governed by $t^{2\nu - \gamma - 1}$ for $\nu>\frac{\gamma}{2}+\frac{1}{2}$. This leads to three distinct cases:

\begin{enumerate}
    \item [B.1.]\textbf{$\nu < \frac{\gamma}{2} + \frac{1}{2}$}: In this regime, both velocity moments remain constant with time, i.e., $\langle|\mathrm{v}|\rangle \propto t^0$ and $\langle \mathrm{v}^2 \rangle \propto t^0$, which results in $M = \frac{1}{2}$ and $L = \frac{1}{2}$.
    
    \item [B.2.]\textbf{$\frac{\gamma}{2} + \frac{1}{2} < \nu < \gamma$}: Here, the first moment remains constant, $\langle|\mathrm{v}|\rangle \propto t^0$, but the second moment scales as $\langle \mathrm{v}^2 \rangle \propto t^{2\nu - \gamma - 1}$, leading to $M = \frac{1}{2}$ and $L = \nu - \frac{\gamma}{2}$.
    
    \item [B.3.] \textbf{$\gamma < \nu < \frac{\gamma}{2} + \eta$}: In this region, both velocity moments grow with time: $\langle|\mathrm{v}|\rangle \propto t^{\nu - \gamma}$ and $\langle \mathrm{v}^2 \rangle \propto t^{2\nu - \gamma - 1}$. This yields $M = \nu - \gamma + \frac{1}{2}$ and $L = \frac{\gamma}{2}$.
\end{enumerate}


For both $\gamma < 1$ and $\gamma > 1$, the $L$ and $M$ exponents of VGLWs coincide with those of GLWs, except in the non-scaling regime, which occurs when $2\nu \geq \gamma + 2\eta$. In this regime, the first and second moments of the speed diverge, and the scaling relations no longer hold.  Figures~\ref{Moses_exponent_variable} and~\ref{Noah_exponent_variable} present the phase portraits of VGLWs, summarizing the different dynamical regimes based on the Moses and Noah exponents. The non-scaling regime appears above the dashed line, which represents the boundary of the scaling domain. In both figures, the dashed line corresponds to the case $\eta = 1$.  
\section{J for VGLWs}

As discussed in Section~\ref{3rd_section}, the Joseph exponent $J$ is obtained from the
scaling of the TAMSD, defined in Eq.~\ref{TAMSD_J}.  
For GLWs, this calculation was carried out in~\cite{albers2022nonergodicity}.  
Here we extend the analysis to VGLWs and show that $J$ is identical to the GLW case; 
in particular, it remains independent of the parameter $\eta$.

The TAMSD of VGLWs is related to the velocity autocorrelation function through the
scaling Green--Kubo relation~\cite{meyer2017scale}:
\begin{equation}
\label{tamsd_green_kubo}
\left\langle \overline{x^2(t,\Delta)} \right\rangle
\approx 
\frac{2}{t}
\int_0^t dt_0
\int_0^\Delta dt_1
\int_0^\Delta dt_2\;
\langle \mathrm{v}(t_1+t_0)\,\mathrm{v}(t_2+t_0) \rangle 
\end{equation}
valid in the limit $t \gg \Delta$.  
The Joseph exponent $J$ is extracted from the $\Delta$-dependence of the above
expression, according to the scaling form in Eq.~\ref{J_scaling}.

For VGLWs, the velocity autocorrelation function (VCF),  
$C(t,t+\Delta)=\langle \mathrm{v}(t)\mathrm{v}(t+\Delta)\rangle$,  
admits the scaling form provided by the generalized Green--Kubo framework
\cite{meyer2017scale},
\begin{equation}
C(t,t+\Delta) 
\simeq 
C_0\, t^{q-2} \,
\phi\!\left( \frac{\Delta}{t} \right),
\end{equation}
where $C_0>0$ is a constant, $q>1$ is a scaling exponent, and  
$\phi(z)$ is a universal scaling function that encodes the dependence on the ratio
$\Delta/t$.  
The scaling function satisfies the asymptotic bounds~\cite{meyer2017scale}
\begin{eqnarray}
\phi(z) < c_1\, z^{-\delta_1},
& \qquad & 2-q \le \delta_1 < 1,
\qquad z \to 0,
\\
\phi(z) < c_u\, z^{-\delta_u},
& \qquad & \delta_u > 1-q,
\qquad z \to \infty,
\end{eqnarray}
where $c_1$ and $c_u$ are positive constants.  
These conditions ensure convergence of Eq.~\eqref{tamsd_green_kubo} and allow the
scaling form of the TAMSD to be determined analytically.

Substituting the scaling Green--Kubo form of the VCF into Eq.~\eqref{tamsd_green_kubo}
and performing the resulting integrals yields a TAMSD that scales as
$\Delta^{2J}$, consistent with Eq.~\ref{J_scaling}.  
Thus, the Joseph exponent for VGLWs coincides with the GLWs result, demonstrating that
$J$ is independent of the velocity-shaping parameter $\eta$.

\subsection{Velocity Correlation Function of VGLWs}




To compute the velocity correlation function (VCF) $\langle \mathrm{v}(t)\mathrm{v}(t+\Delta)\rangle$ for the Variable-Speed Generalized L\'evy Walk (VGLW), we follow the analytical framework developed by Godr\`eche and Luck~\cite{godreche2001statistics}.  
Let $C_n(t,\Delta) \equiv \langle \mathrm{v}(t)\mathrm{v}(t+\Delta)\rangle_n$ denote the contribution to the VCF from trajectories that undergo exactly $n$ renewal events in the interval $(0,t)$.  
Summing over all possible $n$ yields the full VCF:
\begin{equation}
\langle \mathrm{v}(t)\mathrm{v}(t+\Delta)\rangle 
= C(t,\Delta) = \sum_{n=0}^{\infty} C_n(t,\Delta).
\end{equation}

A renewal during the interval $(t,\, t+\Delta)$ destroys correlations between $\mathrm{v}(t)$ and $\mathrm{v}(t+\Delta)$.  
Therefore, the only nonzero contribution to the VCF comes from trajectories for which no renewal occurs between $t$ and $t+\Delta$, meaning both observation times lie within the same step.  

A complete derivation of the VCF, including its asymptotic scaling form and dependence on the parameters $(\gamma,\nu,\eta)$, is provided in Appendix~\ref{vcf_calculation}.

\subsection{TAMSD for VGLWs}






A process observed from the beginning of its evolution ($t = 0$) corresponds to a \textit{non-aging} walk.  
In contrast, when the system is initiated at $t=0$ but observed only after a waiting time $t_a > 0$, the process exhibits \textit{aging}.  
For an aging system, the ensemble-averaged MSD (EAMSD) of a VGLW is defined as~\cite{meyer2017scale}
\begin{equation}
\left\langle x_{t_a}^2(t)\right\rangle
= \left\langle \left[ x(t_a+t) - x(t_a) \right]^2 \right\rangle ,
\end{equation}
which can be computed from the velocity correlation function (VCF)~\cite{meyer2017scale,hu2025generalized}.  
The explicit form of this MSD differs in the \emph{weak aging} regime ($t_a \ll t$) and the \emph{strong aging} regime ($t_a \gg t$).  
For VGLWs, the aging MSD has already been obtained in~\cite{bothe2019mean} using a different approach.

The aging ensemble-averaged TAMSD is defined as  
\begin{equation}
\left\langle \overline{x_{t_a}^2(\Delta,t)} \right\rangle
= \frac{1}{t-\Delta}
\int_{t_a}^{t_a+t-\Delta}
\left\langle \left[ x(t_0+\Delta) - x(t_0) \right]^2 \right\rangle dt_0 .
\end{equation}
Its behavior also depends on whether the system is in the weak- or strong-aging regime.

For determining the Joseph exponent $J$, we use the \emph{weak-aging} TAMSD of the VGLW.  
In the limit $t_a \ll t$ with $t_a \to 0$, the aging TAMSD reduces to the non-aging expression in Eq.~\ref{TAMSD_J}.  
We note, however, that the dependence on the lag time $\Delta$ is identical in both the weak- and strong-aging regimes.

Below we list the general forms of the weak-aging MSD for the VGLW.

\begin{figure}
	\centering
	\includegraphics[width=0.65\columnwidth]{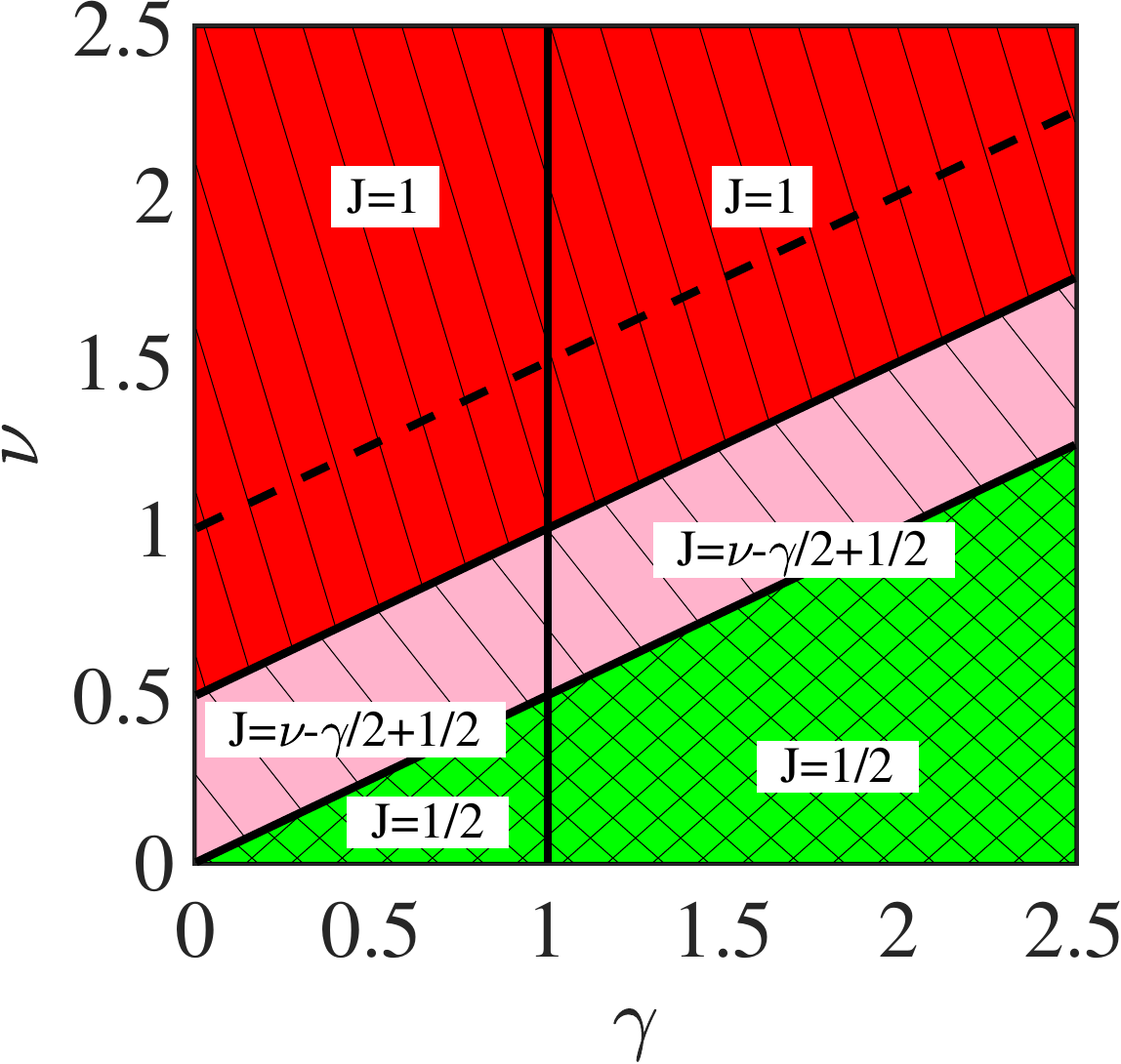}
	\caption{Phase diagram of the Joseph effect for variable speed generalized L\'evy walk .The dotted line marks the onset for infite regime for $\eta=1$}
	{\label{Joseph_exponent_variable}}
\end{figure}

\begin{widetext}
For $\gamma<1$,
\begin{equation}\label{J_gl1}
\left\langle\overline{x^2(t,\Delta)}\right\rangle \propto \begin{cases}\frac{\gamma\eta^2c^2 \Gamma(2 \nu-\gamma-1)}{(2(\nu-\eta)-\gamma)\Gamma(1-\gamma)\Gamma(\gamma)\Gamma(2\nu-\gamma+\eta)} t^{2 \nu-2} \Delta^2, & \gamma / 2+\eta<\nu \\ \frac{2\gamma\eta^2c^2 B(1+\gamma+\eta-2\nu, \eta)}{(2(\nu-\eta)-\gamma)\Gamma(1-\gamma)\Gamma(\gamma)(2\nu-\gamma)(2\nu-\gamma+1)} t^{\gamma-1} \Delta^{1+2 \nu-\gamma}, & \gamma / 2<\nu<\gamma / 2+1 / 2 \\ \frac{2\gamma\eta^2c^2}{(2(\nu-\eta)-\gamma)\Gamma(1-\gamma)\Gamma(\gamma)} \frac{B(1+\gamma+\eta-2\nu, \eta)}{(\gamma-2 \nu)} t^{\gamma-1} \Delta, & \nu<\gamma / 2\end{cases}
\end{equation}
For $\gamma>1$,
\begin{equation}\label{J_gg1}
\left\langle\overline{x^2(t,\Delta)}\right\rangle \propto \begin{cases}\frac{\gamma\eta^2c^2}{(2(\nu-\eta)-\gamma)\langle\tau\rangle} \frac{1}{(2 \nu-\gamma)(2 \nu-\gamma-1)} t^{2 \nu-\gamma-1} \Delta^2, & \gamma / 2+\eta<\nu \\ \frac{2 \gamma\eta^2c^2}{(2(\nu-\eta)-\gamma)\langle\tau\rangle} \frac{B(1+\gamma-2 \nu, \eta)}{(2 \nu-\gamma)(2 \nu-\gamma+1)} \Delta^{1+2 \nu-\gamma}, & \gamma / 2<\nu<\gamma / 2+1 / 2 \\ \frac{2 \gamma\eta^2c^2}{(2(\nu-\eta)-\gamma)\langle\tau\rangle} \frac{B(1+\gamma-2 \nu, \eta)}{(\gamma-2\nu)} \Delta, & \nu<\gamma / 2\end{cases}
\end{equation}
\end{widetext}
Using Eq.\ref{J_gl1} and \ref{J_gg1}, we can write
\begin{equation}
J \propto \begin{cases}1, & \gamma / 2+\eta <\nu \\ (1+2 \nu-\gamma)/2, & \gamma / 2<\nu<\gamma / 2+1 / 2 \\ 1/2, & \nu<\gamma / 2\end{cases}
\end{equation}
The non-scaling regime occurs above the line $2\nu\geq\gamma+2\eta$. Fig. \ref{Joseph_exponent_variable} shows the phase portrait summarizing different regimes based on the Joseph exponent. The dashed line denotes the boundary of scaling regime for $\eta=1$.

\section{$H$ for VGLW}

The Hurst exponent $H$ is determined from the scaling of the MSD, defined in
Eq.~\ref{MSD}. The MSD of VGLWs was analytically derived in~\cite{bothe2019mean};
here we summarize the main results relevant to our analysis.

The propagator \(P(x,t)\), giving the probability density of finding the walker at position
\(x\) at time \(t\), can be expressed as  
\begin{equation}
\label{H_Vglw}
P(x,t)
= \int_{-\infty}^{\infty} \mathrm{d}x'\!
  \int_0^t \mathrm{d}t'\;
  A(x',t')\,
  r\!\left(x - x' \mid t - t'\right),
\end{equation}
where \(A(x',t')\) is the joint density of completing the last full step at \((x',t')\), and
\(r(\cdot)\) is the propagator associated with the remaining (incomplete) portion of the
trajectory.

Taking Fourier transforms of \(P\), \(A\), and \(r\) and expanding for 
\(k \to 0\) yields
\begin{equation}
\begin{aligned}
\widetilde{P}(k,t) &= 1 - \frac{k^2}{2}\,x_2(t) + o(k^2), \\
\widetilde{r}(k,t) &= r_0(t) - \frac{k^2}{2}\,r_2(t) + o(k^2), \\
\widetilde{A}(k,t) &= A_0(t) - \frac{k^2}{2}\,A_2(t) + o(k^2),
\end{aligned}
\end{equation}
where \(x_2(t)\) is the MSD, and \(A_2(t)\), \(r_2(t)\) are the second moments of the
completed and incomplete displacements. Zeroth moments are denoted by
\(A_0(t)=\int A(x,t)\,dx\) and \(r_0(t)=\int r(x,t)\,dx\).

Applying Laplace transforms and the convolution theorem leads to
\begin{equation}
\hat{P}(k,s)
= A_0(s)\,r_0(s)
  - \frac{k^2}{2}\!\left[
      A_0(s)\,r_2(s) + A_2(s)\,r_0(s)
    \right]
  + o(k^2),
\end{equation}
so that the Laplace transform of the MSD is
\begin{equation}
\label{x2_laplace}
\langle x^2(s) \rangle
  = A_0(s)\,r_2(s) + A_2(s)\,r_0(s).
\end{equation}

From the small-$s$ behavior of the above expression, the Hurst exponent follows as  
\begin{equation}
\label{H_g_l_1}
H \propto 
\begin{cases}
\nu, & \gamma/2 < \nu < \gamma/2 + \eta, \\[4pt]
\gamma/2, & \nu < \gamma/2,
\end{cases}
\qquad (\gamma < 1),
\end{equation}
and
\begin{equation}
\label{H_g_g_1}
H \propto 
\begin{cases}
\nu - \gamma/2 + 1/2, 
  & \gamma/2 < \nu < \gamma/2 + \eta, \\[4pt]
1/2, 
  & \nu < \gamma/2,
\end{cases}
\qquad (\gamma > 1).
\end{equation}

These results demonstrate that the Hurst exponent for VGLWs coincides with the
corresponding GLW result. The parameter $\eta$ influences only the prefactors and does
not affect the MSD scaling. A non-scaling (infinite) regime arises whenever
\(2\nu \ge \gamma + 2\eta\).

\begin{figure}
	\centering
	\includegraphics[width=0.65\columnwidth]{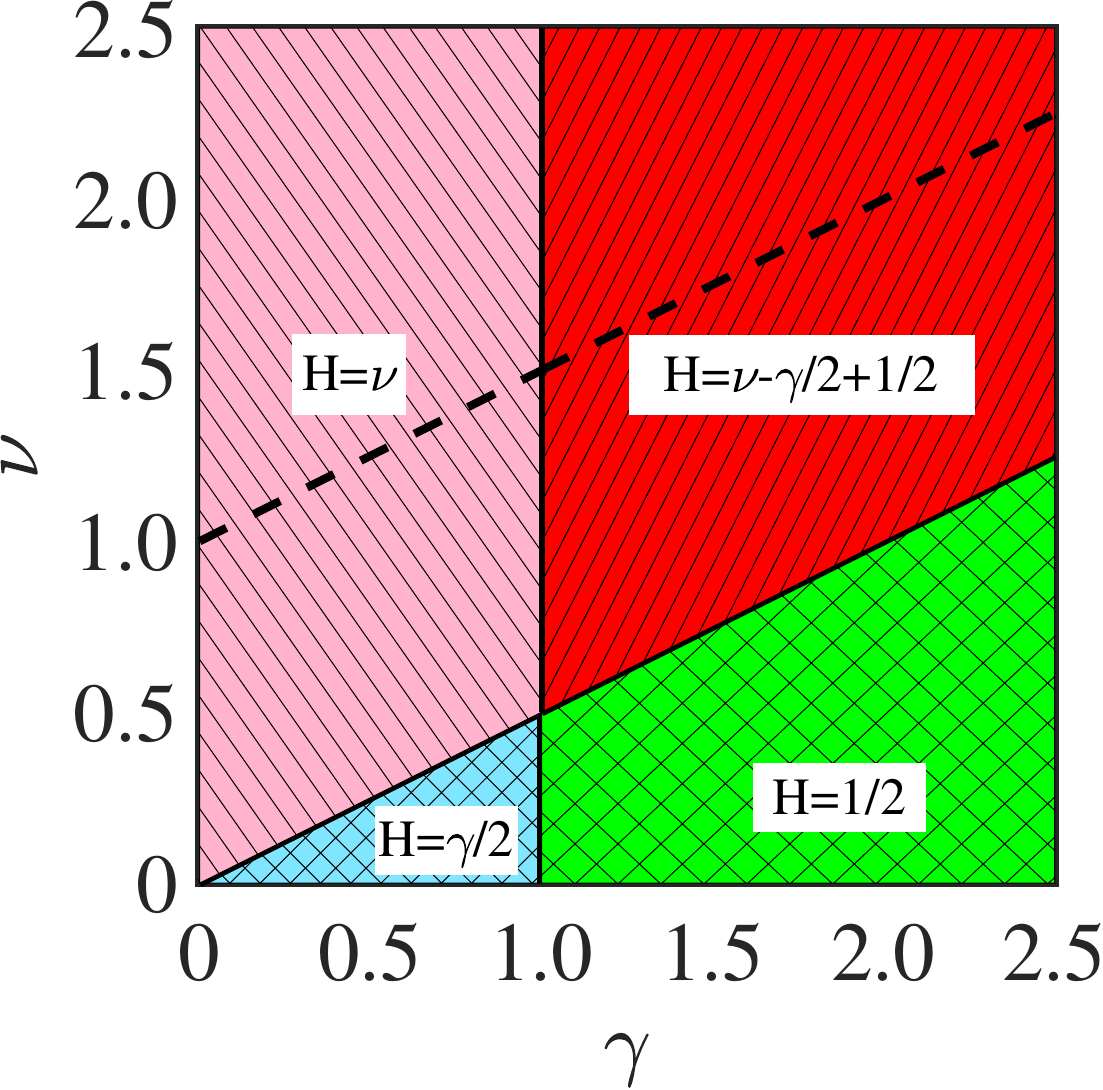}
	\caption{Phase diagram of the Hurst effect for variable speed generalized L\'evy walk .The dotted line marks the onset for infinite regime for $\eta=1$}
	{\label{Hurst_exponent_variable}}
\end{figure}

The resulting dynamical regimes are summarized in
Fig.~\ref{Hurst_exponent_variable}, which shows the phase portrait of $H$; the dashed
line indicates the boundary of the scaling region for $\eta=1$.

\section{Phase Diagram of VGLWS}

\begin{figure}
	\centering
	\includegraphics[width=0.65\columnwidth]{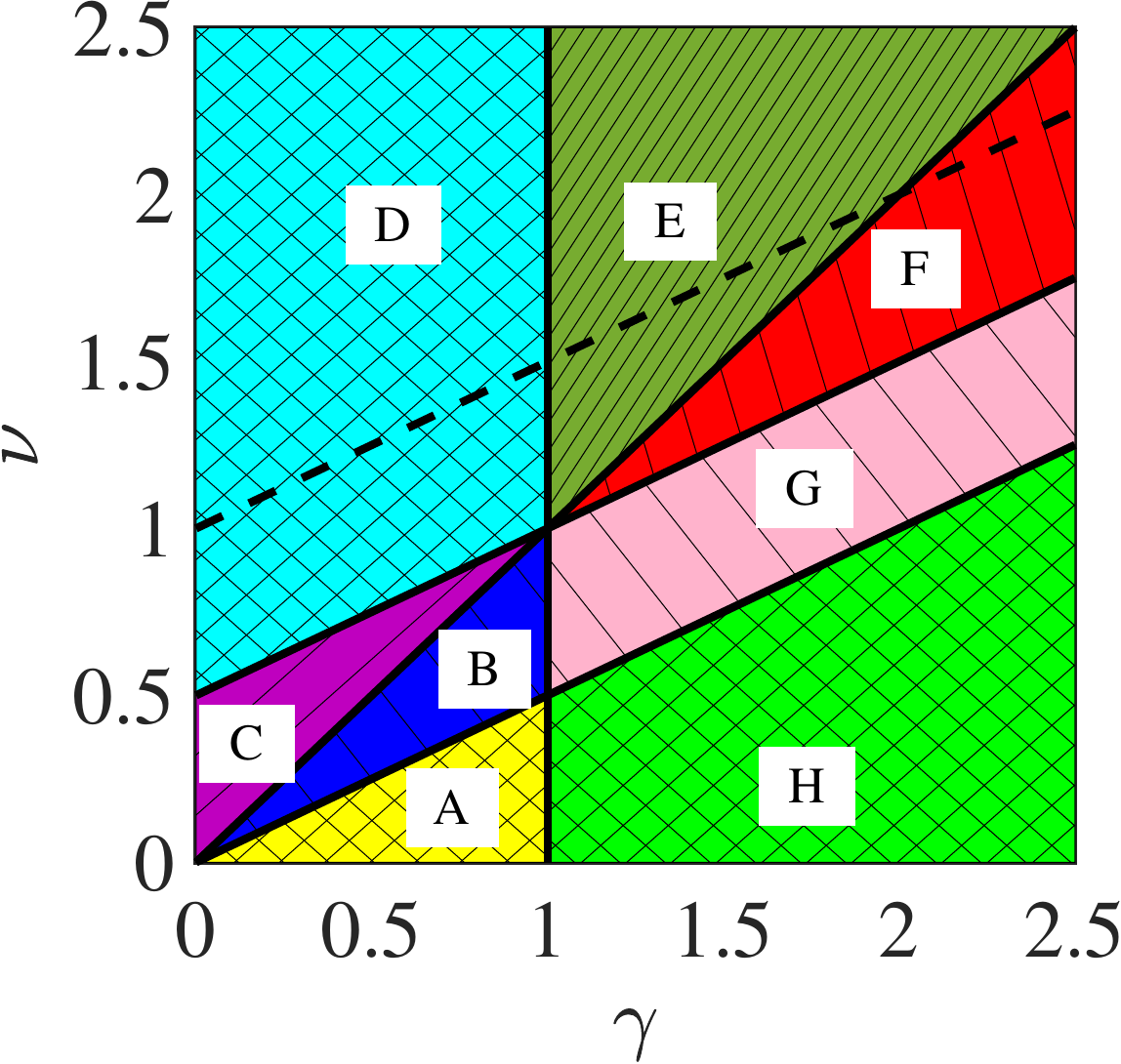}
	\caption{Overall phase diagram of VGLWs based on the full set of scaling
	exponents $(H,J,L,M)$.  
	The dotted line marks the onset of the non-scaling (infinite) regime for $\eta = 1$.}
	\label{Overall_phase_variable}
\end{figure}

The overall phase structure of the VGLW model closely mirrors that of the GLW, even
after introducing the additional parameter $\eta$. The main change arises from the new
boundary of the non-scaling (infinite) regime, which is shifted to the region
\[
2\nu \ge \gamma + 2\eta.
\]
Physically, this extension allows VGLWs to access regions of parameter space that lie
inside the infinite-regime sector of GLWs. A direct consequence is that the Noah
exponent $L$ can exceed unity—an outcome impossible in the classical GLW framework.
The phase portrait for $\gamma < 1$ was previously described in~\cite{aghion2021moses};
here we provide the full VGLW phase diagram, shown in
Fig.~\ref{Overall_phase_variable}.

\noindent
The different dynamical regimes are as follows:

\begin{enumerate}
    \item \textbf{Region A: $\gamma<1$ and $\nu<\gamma/2$}  
    \[
    H=\gamma/2,\quad
    J=1/2,\quad
    L=1-\gamma/2,\quad
    M=\gamma-1/2.
    \]
    No Joseph effect is present (increments are uncorrelated).  
    Anomalous diffusion arises solely from the Moses (aging) and Noah (heavy-tailed)
    effects.

    \item \textbf{Region B: $\gamma<1$ and $\gamma/2<\nu<\gamma$}  
    \[
    H=\nu,\quad
    J=\tfrac{1+2\nu-\gamma}{2},\quad
    L=1-\gamma/2,\quad
    M=\gamma-1/2.
    \]
    All three effects contribute.  
    Joseph correlations are present but not maximal; Moses and Noah effects are
    coupled.

    \item \textbf{Region C: $\gamma<1$ and $\gamma<\nu<\gamma/2+1/2$}  
    \[
    H=\nu,\quad
    J=\tfrac{1+2\nu-\gamma}{2},\quad
    L=1-\nu+\gamma/2,\quad
    M=\nu-1/2.
    \]
    Again, all constitutive effects are active.  
    Joseph correlations are sub-maximal; Noah and Moses remain interdependent.

    \item \textbf{Region D: $\gamma<1$ and $\gamma/2+1/2<\nu<\gamma/2+\eta$}  
    \[
    H=\nu,\quad
    J=1,\quad
    L=1/2,\quad
    M=\nu-1/2.
    \]
    Joseph effect becomes maximal (fully correlated increments).  
    Noah effect disappears (finite variance).  
    Anomalous diffusion is sustained through the Moses (aging) effect.

    \item \textbf{Region E: $\gamma>1$ and $\gamma<\nu<\gamma/2+\eta$}  
    \[
    H=\nu-\gamma/2+1/2,\quad
    J=1,\quad
    L=\gamma/2,\quad
    M=\nu-\gamma+1/2.
    \]
    All three effects contribute.  
    Joseph effect is maximal; increments are heavy-tailed and age with time.

    \item \textbf{Region F: $\gamma>1$ and $\gamma/2+1/2<\nu<\gamma$}  
    \[
    H=\nu-\gamma/2+1/2,\quad
    J=1,\quad
    L=\nu-\gamma/2,\quad
    M=1/2.
    \]
    Joseph effect is maximal; Moses effect is absent (stationary increment
    distribution).  
    Anomalous diffusion arises from Joseph correlations and the Noah effect.

    \item \textbf{Region G: $\gamma>1$ and $\gamma/2<\nu<\gamma/2+1/2$}  
    \[
    H=\nu-\gamma/2+1/2,\quad
    J=\nu-\gamma/2+1/2,\quad
    L=1/2,\quad
    M=1/2.
    \]
    Only the Joseph effect contributes, and it is not maximal.  
    Noah and Moses effects are absent.

    \item \textbf{Region H: $\gamma>1$ and $\nu<\gamma/2$}  
    \[
    H=1/2,\quad
    J=1/2,\quad
    L=1/2,\quad
    M=1/2.
    \]
    All three constitutive effects are absent.  
    Diffusion is normal (MSD $\propto t$).

    \item \textbf{Infinity (Non-scaling) Region:}  
    The scaling relations break down for  
    \[
    2\nu \ge \gamma + 2\eta,
    \]
    and the decomposition into $(H,J,L,M)$ is no longer valid.
\end{enumerate}

Fig.~\ref{Overall_phase_variable} summarizes these regimes across the full VGLW
parameter space for general $\eta$. The dashed curve illustrates the scaling boundary for
$\eta=1$.

\begin{figure}[]
\centering
\includegraphics[width=0.7\columnwidth]{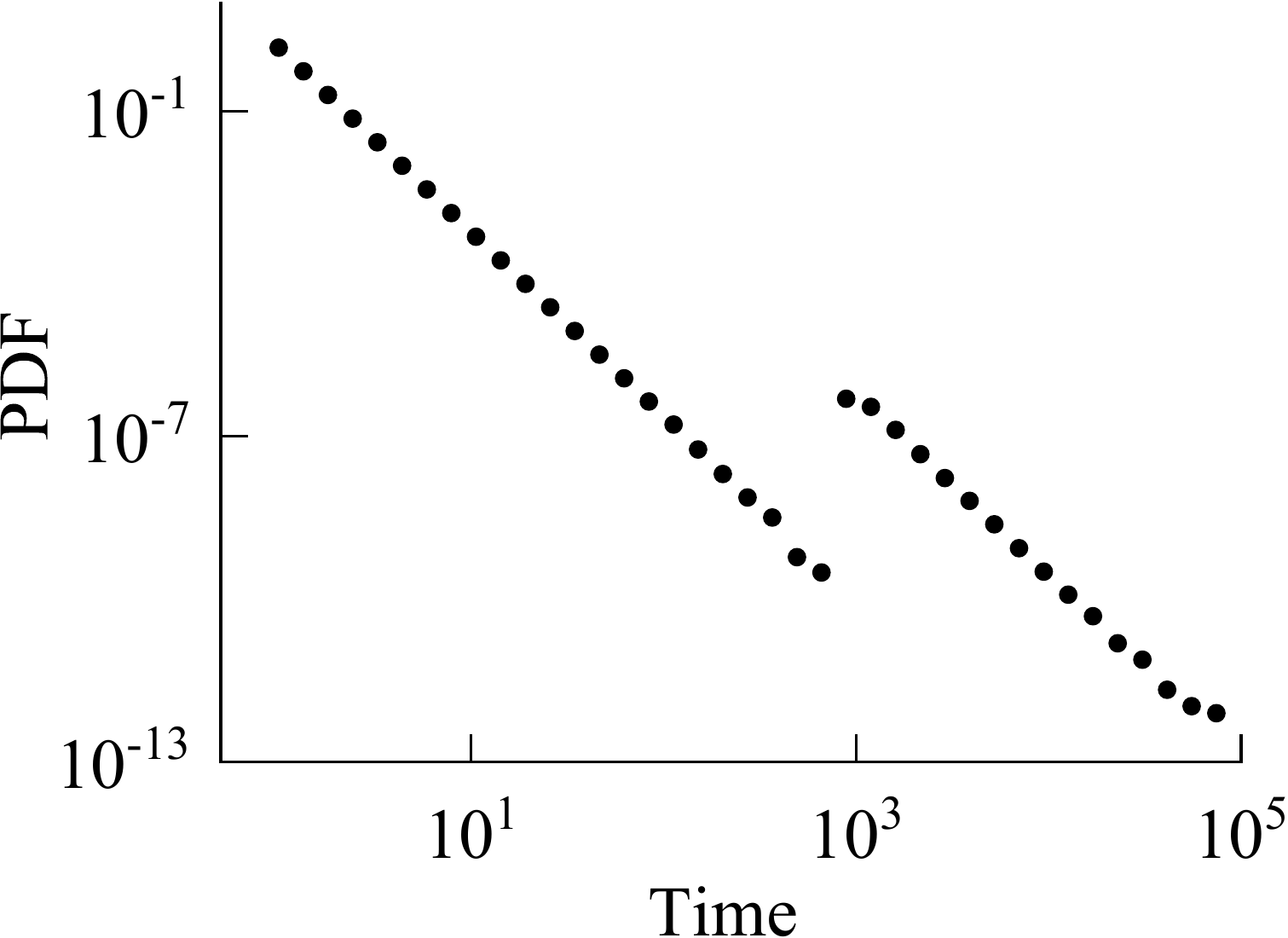}
\caption{Probability density function of time distribution with importance sampling for parameter  $p=4.375\times10^{-4}$ and $t^*=1000$. The points represent logarithmically binned data.}
{\label{Importance_sampling}}
\end{figure}

\section{Numerical Conformations of L$>$1}

\begin{figure}[]{\label{L_greater_than_1}}	
\centering
\includegraphics[width=0.7\linewidth]{"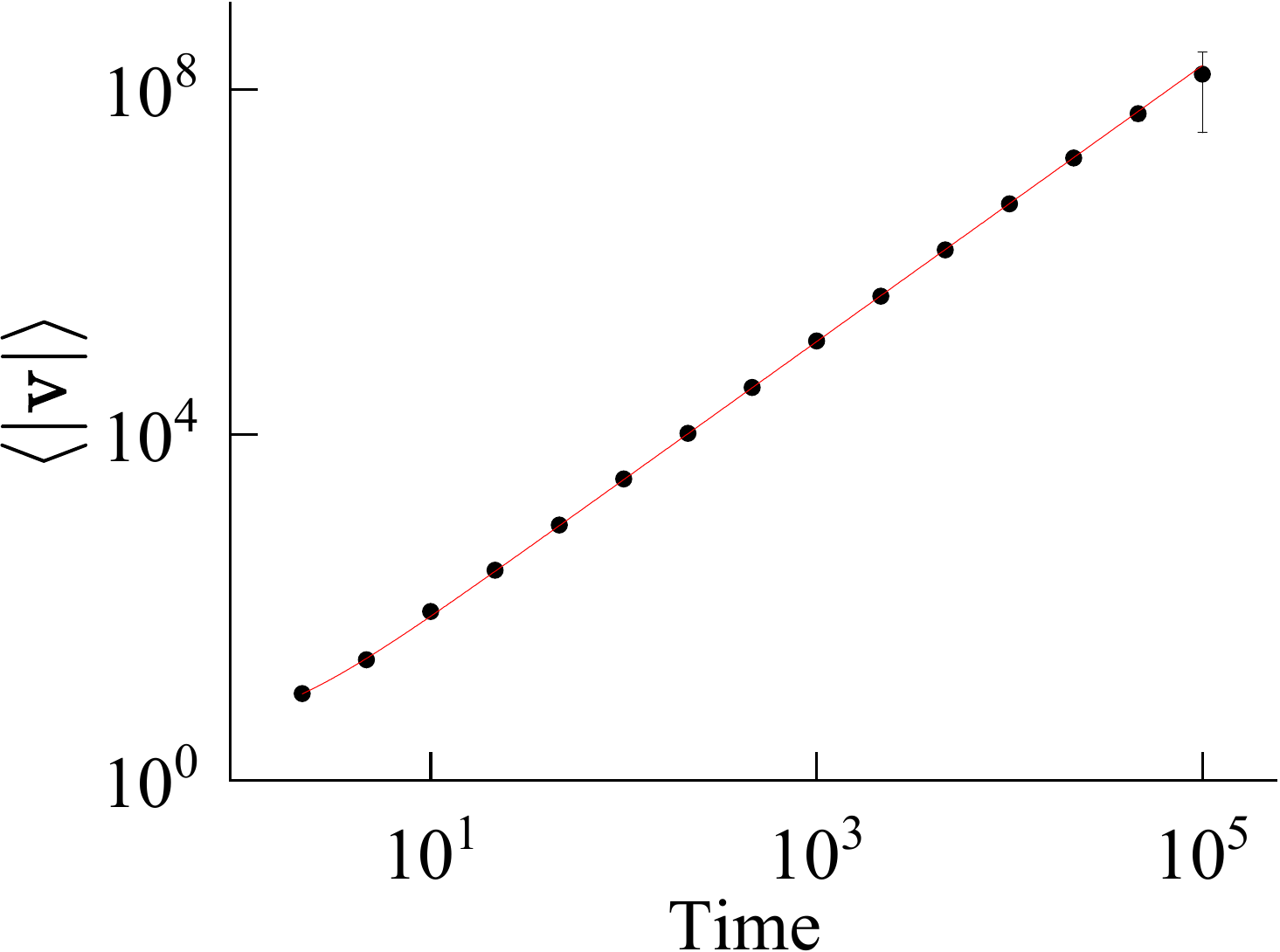"}
\\{\footnotesize (a)}\\
\includegraphics[width=0.7\linewidth]{"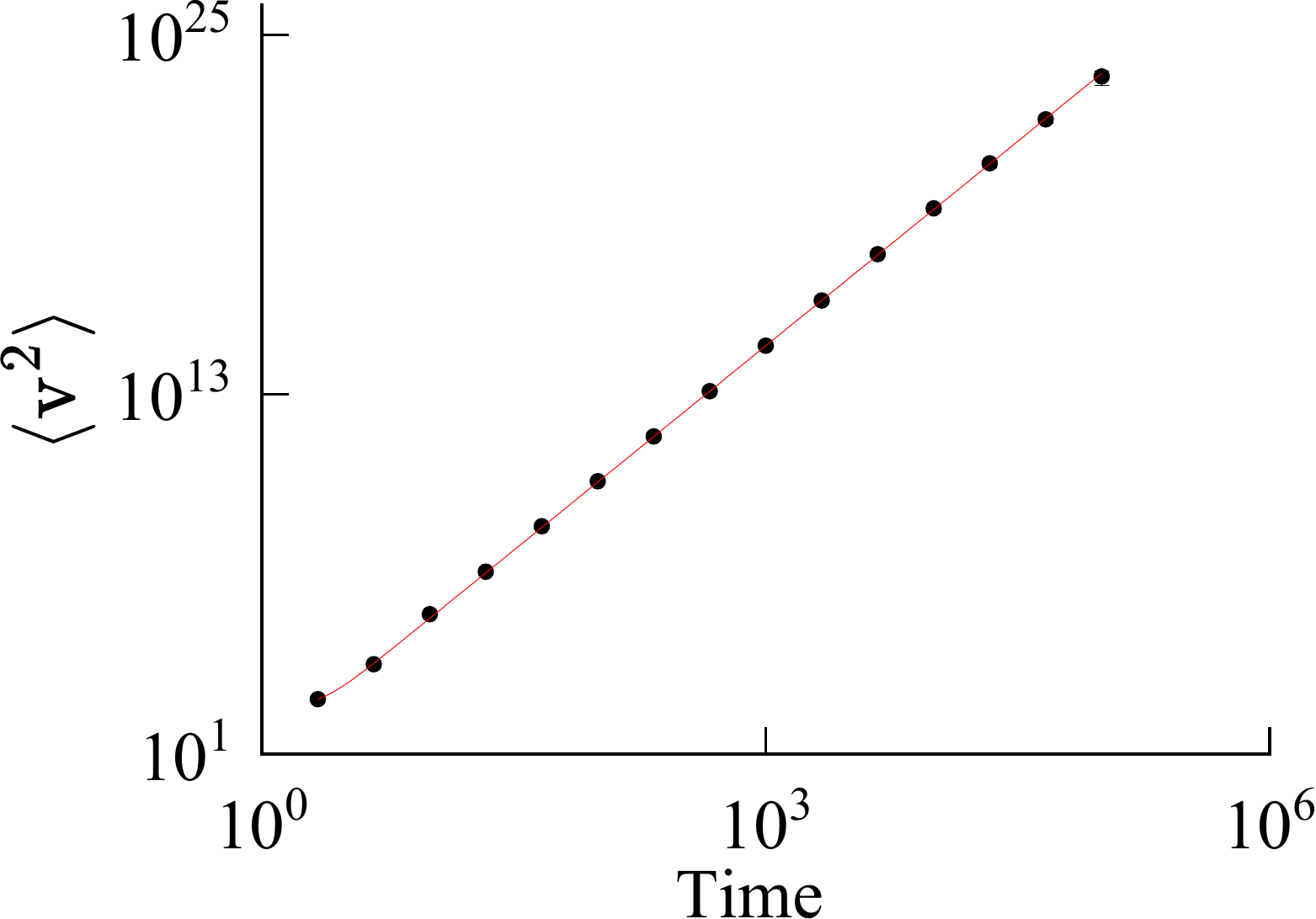"}\\ {\footnotesize (b)}
	\caption{Numerical simulation results at parameters $\gamma = 2.4$, $\nu = 4.0$, and $\eta = 3.0$. (a) Ensemble-averaged absolute velocity. The red line is a fitted straight line with slope $1.60\pm0.0076$. Our analytically predicted value is 1.6. (b) Ensemble-averaged square velocity. The red line is a fitted line with slope $4.55\pm0.01$. Our analytically predicted value is 4.6. $10^7$ random walks were used to obtain these ensemble averages. From these numerical results using Eq.~\ref{Moses_equation} and Eq.~\ref{Noah_equation}, $M=2.10\pm0.0076$ and $L=1.18\pm0.0176$. The analytically predicted values are $M=2.1$ and $L=1.2$.}
	{\label{L_greater_than_1}}
\end{figure}


To confirm our analytical prediction that the Noah exponent $L$ can exceed unity, we have numerically computed the first and second moments of the absolute velocity, $\langle |\mathrm{v}| \rangle$ and $\langle \mathrm{v}^2 \rangle$, for the parameter set $\gamma = 2.4$, $\nu = 4.0$, and $\eta = 3.0$. These parameters correspond to a point in phase~E in Fig.~\ref{Overall_phase_variable}, where our analytical framework predicts $L > 1$.  At such a large value of $\gamma$, it becomes challenging to achieve good statistical sampling of steps with durations in the heavy tail of the distribution given by Eq.~\ref{original_time_distribution}. To address this issue, we employed a biased (importance) sampling technique~\cite{albers2022nonergodicity}. Instead of drawing step durations directly from $\psi(t)$ in Eq.~\ref{original_time_distribution}, we sampled them from a modified distribution $\widetilde{\psi}(t)$ defined as  
\begin{eqnarray}
\label{importance_sampling}
\widetilde{\psi}(t) &=& (1 - p)\,\psi(t) + p\,\psi(t \mid t > t^*), \\
\psi(t \mid t > t^*) &=& \Theta(t - t^*)\,\frac{\psi(t)}{\int_{t^*}^{\infty} \psi(t)\, \mathrm{d}t}.
\end{eqnarray}
Figure~\ref{Importance_sampling} shows $\widetilde{\psi}(t)$ for $\gamma = 2.4$ on a log–log scale with parameters $p = 0.0004375$ and $t^* = 1000$. 



Because the sampling is performed using $\widetilde{\psi}(t)$ rather than $\psi(t)$, the results must be reweighted to eliminate the sampling bias. The weight assigned to each step in a trajectory depends on the duration of that step, $t$, and is given by  
\begin{equation}
L_{\mathrm{step}}(t) = \frac{\psi(t)}{\widetilde{\psi}(t)} =
\begin{cases}
\dfrac{1}{1 - p}, & t < t^*, \\[6pt]
\dfrac{1}{1 - p + p / \lambda}, & t > t^*,
\end{cases}
\end{equation}
where  
\begin{equation}
\lambda = \int_{t^*}^{\infty} \psi(t)\, \mathrm{d}t.
\end{equation}
The total weight associated with an entire walk is then  
\begin{equation}
L_{\mathrm{walk}} = 
\left(\frac{1}{1 - p}\right)^{N_{t < t^*}}
\left(\frac{1}{1 - p + p / \lambda}\right)^{N_{t > t^*}},
\end{equation}
where $N_{t < t^*}$ and $N_{t > t^*}$ denote, respectively, the number of steps in the walk with durations shorter and longer than $t^*$.  


Using biased sampling with parameters $p = 4.375 \times 10^{-4}$ and $t^* = 10^{3}$, we obtained the results shown in Fig.~\ref{L_greater_than_1} for $\langle |\mathrm{v}| \rangle$ and $\langle \mathrm{v}^2 \rangle$ at $\gamma = 2.4$, $\nu = 4.0$, and $\eta = 3.0$. From these numerical simulations, we find that under these parameter values, the Noah exponent is $L = 1.18 \pm 0.0176$ to 2$\sigma$ error, in rough agreement with the analytical prediction of $L = 1.2$. 
\section{Discussion and Conclusion}

VGLWs provide a unified framework that encompasses many previously studied
continuous-time random walk (CTRW) models~\cite{supp}.  
By tuning the parameter $\eta$, the model interpolates smoothly between a number of
important limiting cases.  
For $\eta = 1$, VGLWs reduce to GLWs~\cite{albers2018exact},
and for $\nu = \eta = 1$, they further collapse to the standard L\'evy walk
model~\cite{zaburdaev2015levy,shlesinger1987levy}.  
In the limits $\eta \to 0$ and $\eta \to \infty$~\cite{bothe2019mean}, the dynamics
correspond to the wait--jump and jump--wait L\'evy flight processes, respectively.
When $\nu = \eta$, the model becomes equivalent to the Drude-type dynamics in which the
velocity depends only on the elapsed time of the step~\cite{schulz1997anomalous,benkadda1998chaos}.  

In this work we analyze anomalous diffusion in VGLWs through its decomposition into the
Joseph, Noah, and Moses effects.  
This decomposition reveals nine distinct dynamical phases, each governed by a different
combination of these three constitutive mechanisms.  
Although the model contains three control parameters, the role of $\eta$ is limited:
it affects only the position of the boundary of the non-scaling (infinite) regime, where
the standard diffusive scaling breaks down.

Earlier work~\cite{aghion2021moses} treated only the regime $\gamma < 1$, in which the
mean step duration diverges and the dynamics are dominated by a single long step.
Here, we extend the decomposition to the regime $\gamma > 1$, where the mean step
duration is finite and the statistics emerge from the collective contribution of many
steps.  
For $\gamma < 1$, anomalous diffusion is always generated jointly by the Joseph, Noah,
and Moses effects, producing five distinct dynamical phases.  
In contrast, for $\gamma > 1$ the system may exhibit either anomalous or normal
diffusion.  
Normal diffusion occurs when $\nu < \gamma/2$, where all exponents take the value
$1/2$.  
When $\gamma/2 < \nu < \gamma/2 + 1/2$, anomalous diffusion arises solely from the
Joseph effect.  
For $\gamma/2 + 1/2 < \nu < \gamma$, anomalous diffusion results from the combined
action of the Noah and Joseph effects.  
When $\gamma < \nu < \gamma/2 + \eta$, all three effects act simultaneously.  
Finally, for $\nu > \gamma/2 + \eta/2$, the system enters a non-scaling (infinite)
regime in which the usual scaling laws no longer apply.

A striking feature of the VGLW framework is that, in the range  
\[
\gamma/2 + \eta/2 > \nu > \gamma/2 + 1/2 \quad \text{with} \quad \eta > 1,
\]
the Noah exponent $L$ can exceed unity.  
This contrasts with earlier findings for L\'evy-walk-type processes, where $L \le 1$
was believed to be the upper limit~\cite{chen2017anomalous,aghion2021moses}.  
Our results show that $L$ can surpass $1$ in the regime $\gamma > 1$, demonstrating that
the Noah exponent has no intrinsic upper bound within the VGLW model.

This extension is particularly important because many real-world systems operate in the
$\gamma > 1$ regime.  
Examples include physical, biological, and geological transport processes such as
chaotic advection, groundwater flow, molecular motion, and animal foraging
dynamics~\cite{del2000chaotic,del1998asymmetric,zaburdaev2011perturbation,sims2008scaling,%
berkowitz2016measurements,cortis2004anomalous,levy2003measurement,bijeljic2011signature}.  
Our results enable the identification of which constitutive effect—Joseph, Noah, or
Moses—dominates the observed anomalous behavior in such systems.  
The resulting framework therefore captures the essential physics of VGLW-type
transport and provides deeper insight into the mechanisms driving complex anomalous
diffusion phenomena.

\appendix
\section{The velocity propagator of VGLW}{\label{Velocity_propagator_vglw_definition}}

In this section, we derive the velocity propagator of VGLWs,
$p(\mathrm{v},t)$, defined as the probability density of observing a
walker with velocity $\mathrm{v}$ at time $t$.
Unlike GLWs, the velocity in a VGLW is not constant within a step; instead,
it depends deterministically on both the elapsed time $t'$ since the
beginning of the step and the total step duration $\tau$.

To compute $p(\mathrm{v},t)$, we first introduce the conditional propagator
$p(\mathrm{v},t,t')$, which gives the probability density of observing
velocity $\mathrm{v}$ at time $t$, conditioned on the walker having spent
an elapsed time $t'$ in the current step. The full propagator is obtained by
integrating over all admissible $t'$.

Using the step-duration distribution in
Eq.~\ref{original_time_distribution} and performing a change of variables
from $\tau$ to $\mathrm{v}$, we obtain the conditional velocity distribution
at fixed elapsed time $t'$,
\begin{equation}
\begin{aligned}
\chi(\mathrm{v},t')
&= \psi(\tau)\left|\frac{d\tau}{d\mathrm{v}}\right| \\
&= \frac{\gamma \tau_0^\gamma (\eta c)^{\gamma/(\nu-\eta)}}{\nu-\eta}
\, t'^{\frac{(\eta-1)\gamma}{\nu-\eta}}
\, |\mathrm{v}|^{-\frac{\gamma}{\nu-\eta}-1}.
\end{aligned}
\end{equation}

For fixed elapsed time $t'$, the joint distribution of velocity and step
duration for an incomplete step is
\begin{equation}
\label{eq:joint_phi}
\phi(\mathrm{v},\tau,t')
=
\chi(\mathrm{v},t')\,
\delta\!\left[
\left(\frac{|\mathrm{v}|}{\eta c t'^{\eta-1}}\right)^{\frac{1}{\nu-\eta}}
- \tau
\right]
\Theta(\tau - t').
\end{equation}
Because the dynamics are symmetric with respect to direction,
$\phi(\mathrm{v},\tau,t')=\phi(-\mathrm{v},\tau,t')$.

The escape probability $W(\mathrm{v},t,t')$, defined as the probability that
a walker with elapsed time $t'$ continues the current step beyond time $t$,
is then
\begin{equation}
\begin{aligned}
W(\mathrm{v},t,t')
&= \int_t^\infty \phi(\mathrm{v},\tau,t')\, d\tau \\
&= \chi(\mathrm{v},t')\,
\Theta\!\left[
\left(\frac{|\mathrm{v}|}{\eta c t'^{\eta-1}}\right)^{\frac{1}{\nu-\eta}}
- t
\right]
\Theta(t - t').
\end{aligned}
\end{equation}

The renewal density $R(t)$, representing the probability density for the
initiation of a new step at time $t$, is identical to that of GLWs since it
depends only on the step-duration distribution. Following
Ref.~\cite{akimoto2020infinite},
\begin{equation}
\label{eq:renewal_event}
R(t) = \mathcal{L}^{-1}
\!\left[\frac{1}{1-\widetilde{\psi}(z)}\right],
\end{equation}
where $\widetilde{\psi}(z)$ is the Laplace transform of $\psi(t)$.

The conditional propagator corresponds to a renewal at time $t-t'$ followed
by a step that survives at least until time $t$. It is therefore given by
\begin{equation}
\label{eq:pvttprime}
\begin{aligned}
p(\mathrm{v},t,t')
&= R(t-t')\, W(\mathrm{v},t',t') \\
&= R(t-t')\, \chi(\mathrm{v},t')\,
\Theta\!\left[
\left(\frac{|\mathrm{v}|}{\eta c t'^{\eta-1}}\right)^{\frac{1}{\nu-\eta}}
- t'
\right]\Theta(t-t') \\
&\approx
\frac{\gamma \tau_0^\gamma (\eta c)^{\gamma/(\nu-\eta)}}{\nu-\eta}
\, |\mathrm{v}|^{-\frac{\gamma}{\nu-\eta}-1}
\, R(t-t')\,
t'^{\frac{(\eta-1)\gamma}{\nu-\eta}}
\\
&\quad \times
\Theta\!\left[
\left(\frac{|\mathrm{v}|}{\eta c t'^{\eta-1}}\right)^{\frac{1}{\nu-\eta}}
- t'
\right]\Theta(t-t').
\end{aligned}
\end{equation}

The full velocity propagator is obtained by integrating over all elapsed
times,
\begin{equation}
p(\mathrm{v},t) = \int_0^t p(\mathrm{v},t,t')\, dt'.
\end{equation}

The above construction is valid only when $\nu \neq \eta$.
In the special case $\nu=\eta$, the velocity depends solely on the elapsed
time,
\begin{equation}
\mathrm{v}_{\nu,\eta} = \pm \eta c\, t'^{\eta-1}.
\end{equation}
Since the velocity is independent of the total step duration $\tau$, all
steps with duration exceeding $t'$ contribute equally.
The conditional propagator then reduces to
\begin{equation}
\begin{aligned}
p(\mathrm{v},t,t')
=
R(t-t')\,
\delta\!\left(|\mathrm{v}|-\eta c t'^{\eta-1}\right)
\Theta(t-t')\!
\int_{t'}^\infty \psi(\tau)\, d\tau .
\end{aligned}
\end{equation}
The full propagator $p(\mathrm{v},t)$ is again obtained by integration over
$t'$.

In the following sections, we analyze the structure and asymptotic behavior
of $p(\mathrm{v},t)$ resulting from these expressions. The detalied calculations are shown in appendix.~\ref{Detailed_calculation_Propagator}.

\subsection{$\gamma<1$}

For $\gamma<1$, the mean step duration diverges and the renewal rate
$R(t)$ defined in Eq.~\ref{eq:renewal_event} scales as
$R(t)\propto t^{\gamma-1}$.
In this regime, the velocity propagator exhibits universal scaling that is
independent of the microscopic time scale $t_0$.

\subsubsection{$\nu<1$}

We first consider the case $\nu<1$ and study how the structure of
$p(\mathrm{v},t)$ evolves as the parameter $\eta$ is varied.
When $\eta=1$, the model reduces to a GLW, and the corresponding propagator
for $\gamma<1$ and $\nu<1$ is discussed in~\cite{akimoto2020infinite}.

\paragraph{Case $\nu<\eta$}

For fixed $t$, the velocity propagator takes the form
\begin{equation}
p(\mathrm{v},t)\approx
\begin{cases}
k_0\, \mathrm{v}^{-\frac{\gamma}{\nu-\eta}-1}\,
t^{\gamma+\frac{(\eta-1)\gamma}{\nu-\eta}},
& \mathrm{v}<\mathrm{v}_c(t),\\[6pt]
k_1\, t^{\gamma-1}\,
\mathrm{v}^{-\frac{\gamma}{\nu-1}-1+\frac{1}{\nu-1}},
& \mathrm{v}>\mathrm{v}_c(t),
\end{cases}
\end{equation}
where $k_0$ and $k_1$ are constants given in Eq.~\ref{value_of_constants}, and
the critical velocity is
\[
\mathrm{v}_c(t)=\eta c\, t^{\nu-1}.
\]
The propagator is piecewise continuous: below $\mathrm{v}_c(t)$ the scaling
depends explicitly on $\eta$, while above $\mathrm{v}_c(t)$ the scaling is
$\eta$-independent and originates from the infinite invariant density.

\paragraph{Case $\nu=\eta$}

As $\eta$ is decreased toward $\nu$, the propagator becomes
\begin{equation}
p(\mathrm{v},t)\approx
\begin{cases}
0,
& \mathrm{v}<\mathrm{v}_c(t),\\[6pt]
\displaystyle
\frac{t^{\gamma-1}}{(\eta-1)\gamma\,|\Gamma(1-\gamma)|\Gamma(\gamma)}
\\\times\left(\frac{\mathrm{v}}{\eta c}\right)^{-\frac{\gamma}{\eta-1}
+\frac{1}{\eta-1}-1},
& \mathrm{v}>\mathrm{v}_c(t).
\end{cases}
\end{equation}
In this case, $p(\mathrm{v},t)$ is entirely governed by the
infinite invariant density.

\paragraph{Case $\nu>\eta$}

For $\nu>\eta$, the propagator reads
\begin{equation}
p(\mathrm{v},t)\approx
\begin{cases}
0,
& \mathrm{v}<\mathrm{v}_c(t),\\[6pt]
k_0\, \mathrm{v}^{-\frac{\gamma}{\nu-\eta}-1}\,
t^{\gamma+\frac{(\eta-1)\gamma}{\nu-\eta}}
\\- k_1\, t^{\gamma-1}\,
\mathrm{v}^{-\frac{\gamma}{\nu-1}-1+\frac{1}{\nu-1}},
& \mathrm{v}>\mathrm{v}_c(t).
\end{cases}
\end{equation}
Here, the propagator is continuous above $\mathrm{v}_c(t)$.
The first term determines the $\eta$-dependent scaling, while the second
term reflects the contribution from the infinite invariant density.

\subsubsection{$\nu>1$}

We now consider $\gamma<1$ with $\nu>1$, again varying $\eta$.

\paragraph{Case $\nu<\eta$}

For fixed $t$, the propagator is
\begin{equation}
p(\mathrm{v},t)\approx
\begin{cases}
k_0\, \mathrm{v}^{-\frac{\gamma}{\nu-\eta}-1}\,
t^{\gamma+\frac{(\eta-1)\gamma}{\nu-\eta}}
\\- k_1\, t^{\gamma-1}\,
\mathrm{v}^{-\frac{\gamma}{\nu-1}-1+\frac{1}{\nu-1}},
& \mathrm{v}<\mathrm{v}_c(t),\\[6pt]
0,
& \mathrm{v}>\mathrm{v}_c(t).
\end{cases}
\end{equation}
The propagator is continuous below $\mathrm{v}_c(t)$.
The $\eta$-dependent term governs the scaling, while the second term again
originates from the infinite invariant density.

\paragraph{Case $\nu=\eta$}

At $\nu=\eta$, the propagator simplifies to
\begin{equation}
p(\mathrm{v},t)\approx
\begin{cases}
\displaystyle
\frac{t^{\gamma-1}}{(\eta-1)\gamma\,|\Gamma(1-\gamma)|\Gamma(\gamma)}
\\\times\left(\frac{\mathrm{v}}{\eta c}\right)^{-\frac{\gamma}{\eta-1}
+\frac{1}{\eta-1}-1},
& \mathrm{v}<\mathrm{v}_c(t),\\[6pt]
0,
& \mathrm{v}>\mathrm{v}_c(t).
\end{cases}
\end{equation}
As before, the propagator is continuous and dominated by the infinite
invariant density.

\paragraph{Case $\nu>\eta$}

Finally, for $\nu>\eta$ we obtain
\begin{equation}
p(\mathrm{v},t)\approx
\begin{cases}
k_1\, t^{\gamma-1}\,
\mathrm{v}^{-\frac{\gamma}{\nu-1}-1+\frac{1}{\nu-1}},
& \mathrm{v}<\mathrm{v}_c(t),\\[6pt]
k_0\, \mathrm{v}^{-\frac{\gamma}{\nu-\eta}-1}\,
t^{\gamma+\frac{(\eta-1)\gamma}{\nu-\eta}},
& \mathrm{v}>\mathrm{v}_c(t).
\end{cases}
\end{equation}
The propagator is piecewise continuous: below $\mathrm{v}_c(t)$ the scaling
is $\eta$-independent and controlled by the infinite invariant density,
whereas above $\mathrm{v}_c(t)$ the scaling depends explicitly on $\eta$.

\begin{figure}
	\centering
	\includegraphics[width=0.75\columnwidth]{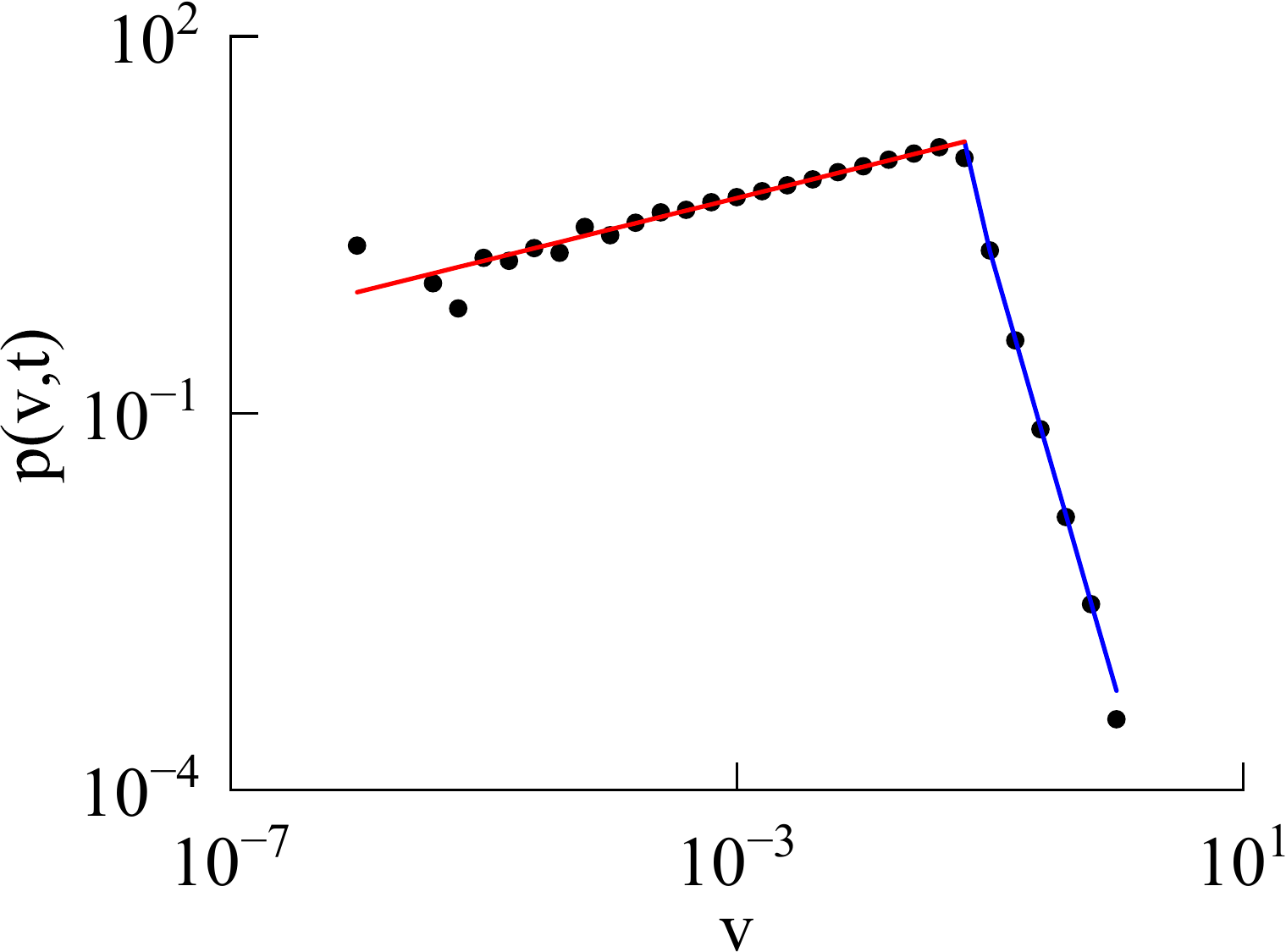}\\
	{\footnotesize (a)}\\
	\includegraphics[width=0.75\columnwidth]{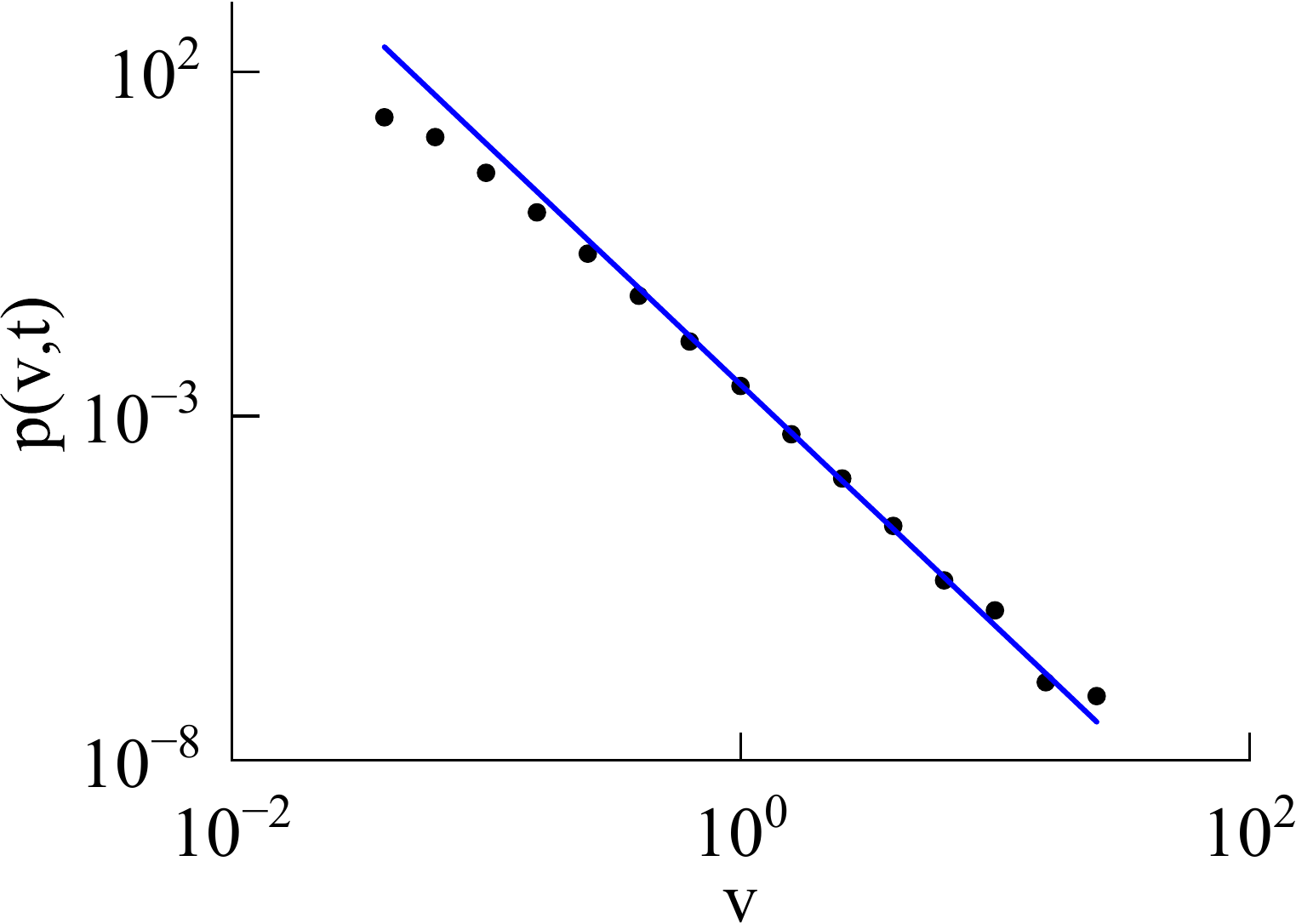}\\
	{\footnotesize (b)}\\
	\includegraphics[width=0.75\columnwidth]{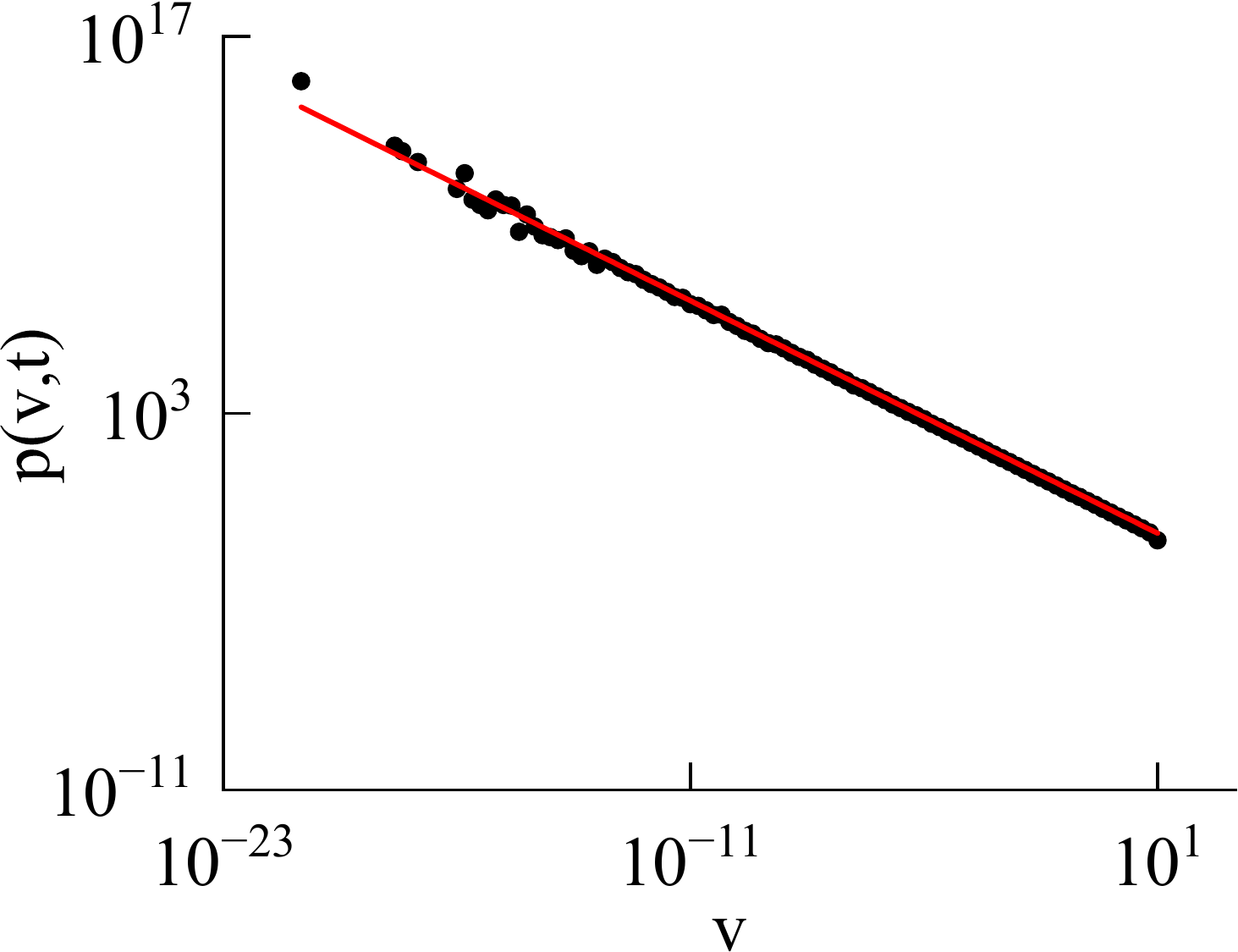}\\
	{\footnotesize (c)}\\
	\includegraphics[width=0.75\columnwidth]{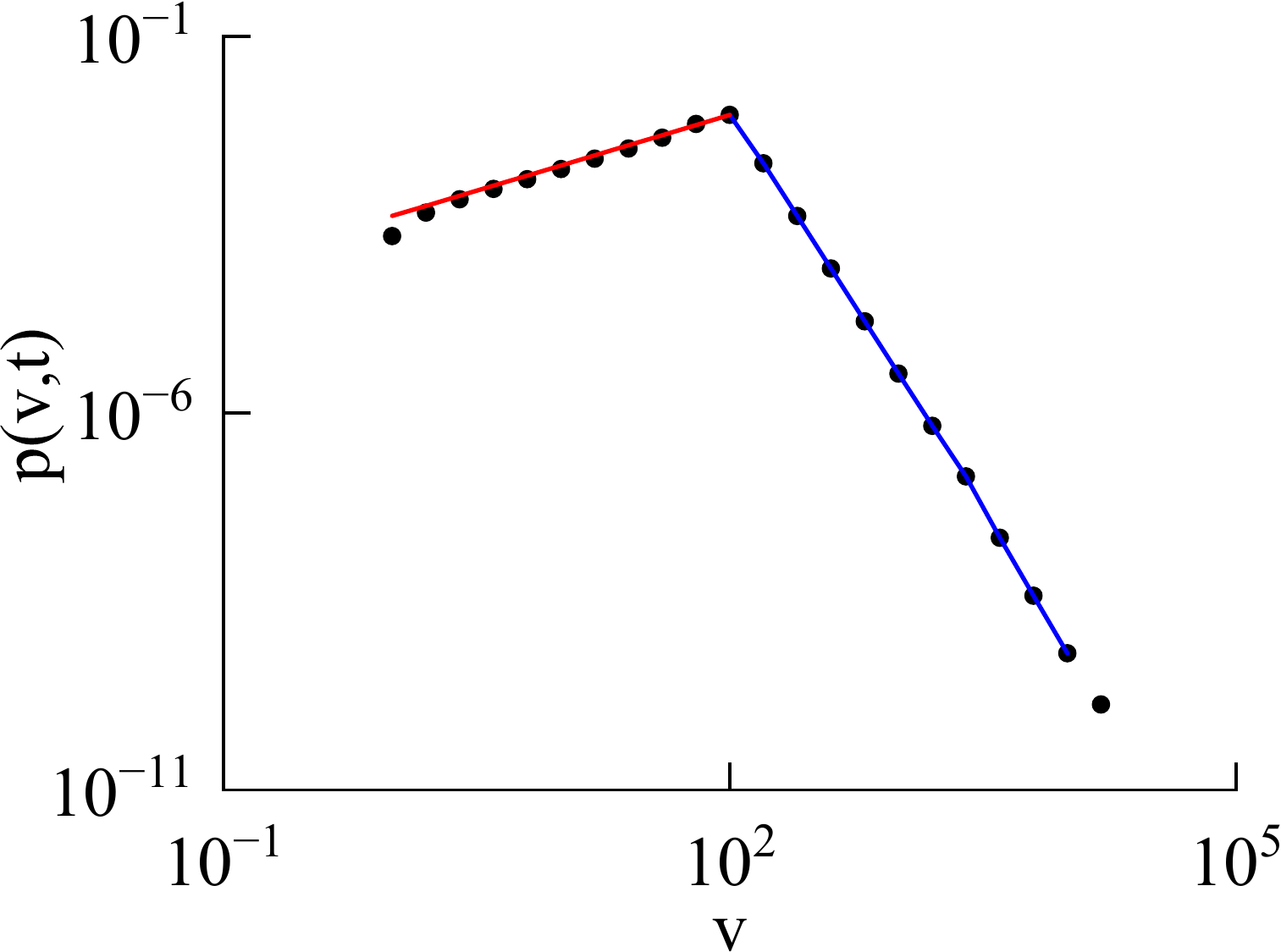}\\
	{\footnotesize (d)}
	\caption{Velocity propagator $p(\mathrm{v},t)$ for $\gamma=0.5$ at fixed
	$t=10^6$, obtained from $10^6$ realizations.
	(a) $\nu<\eta$ and $\nu<1$ $(\nu=0.8,\eta=1.2)$;
	(b) $\nu>\eta$ and $\nu<1$ $(\nu=0.8,\eta=0.6)$;
	(c) $\nu<\eta$ and $\nu>1$ $(\nu=1.1,\eta=2.9)$;
	(d) $\nu>\eta$ and $\nu>1$ $(\nu=1.3,\eta=1.1)$.The points represent simulation results, while the solid lines denote the analytical predictions. The red and blue line denotes the $p(\mathrm{v},t)$ before and after $\mathrm{v}_c(t)$, respectively.}
	\label{velocity_propagator_vglw}
\end{figure}

Figure~\ref{velocity_propagator_vglw} shows the velocity propagator in
log--log scale for representative parameter choices.
The step-time distribution is
$\psi(\tau)=\gamma\tau^{-1-\gamma}\Theta(\tau\ge1)$.
The numerical results confirm the analytical predictions and illustrate
the universal scaling of $p(\mathrm{v},t)$ for $\gamma<1$.

\subsection{$\gamma>1$}
For $\gamma>1$, the mean step duration is finite and the renewal rate
$R(t)$ defined in Eq.~\ref{eq:renewal_event} approaches a constant,
$R(t)\propto t^{0}$ or independent of time.

\subsubsection{$\nu<1$}
For this particular choice of $\gamma$ and $\nu$, we would slowly tune the parameter $\eta$ to observe the nature of $p(\mathrm{v},t)$. 
We would start with $\nu<\eta$
For fixed $t$, $p(\mathrm{v},t)$ would be  
\begin{equation}
p(\mathrm{v}, t)\approx \begin{cases}\frac{(\gamma-1)\tau_0^{(\gamma-1)} \left(\eta c\right)^{\frac{\gamma-1}{\nu-1}}}{(\eta-1)\gamma+(\nu-\eta)}\mathrm{v}^{\frac{-\gamma}{\nu-\eta}-1}\left(t^{1+\frac{\left(n-1 \right)\gamma}{\nu-\eta}}\right), & \mathrm{v}<\mathrm{v}_c(t) \\ \frac{(\gamma-1)\tau_0^{(\gamma-1)} \left(\eta c\right)^{\frac{\gamma}{\nu-\eta}}}{(\eta-1)\gamma+(\nu-\eta)}\left(t^{0}\mathrm{v}^{\frac{-\gamma}{\nu-1}-1+\frac{1}{\nu-1}}\right), & \mathrm{v}>\mathrm{v}_c(t)\end{cases}
\end{equation}
Where critical velocity $\mathrm{v}_c(t)=\eta c\left(t\right)^{\nu-1}$. For this parameter choice, $p(\mathrm{v}, t)$ is a piece-wise continuous function. Before $\mathrm{v}_{c}(t)$ , the scaling is $\eta$ dependent. After $\mathrm{v}_{c}(t)$.  the scaling is $\eta$ independent.

We would decrease the $\eta$ parameter till $\nu=\eta$
\begin{equation}
p(\mathrm{v}, t)\approx \begin{cases} 0 , &\mathrm{v}<\mathrm{v}_c(t) \\ \frac{(\gamma-1)\tau_0^{\gamma-1}}{\gamma(\eta-1)} t^{0}{\frac{\mathrm{v}}{\eta c}}^{-\frac{\gamma}{\eta-1}+\frac{1}{\eta-1}-1}, & \mathrm{v}>\mathrm{v}_c(t)\end{cases}
\end{equation}
We would further decrease the $\eta$ parameter so that $\nu>\eta$
For fixed $t$, $p(\mathrm{v},t)$ would be  
\begin{equation}
p(\mathrm{v}, t)\approx \begin{cases}0 ,&\mathrm{v}<\mathrm{v}_c(t) \\ \frac{(\gamma-1)\tau_0^{(\gamma-1)} \left(\eta c\right)^{\frac{\gamma}{\nu-\eta}}}{(\eta-1)\gamma+(\nu-\eta)}\left(\mathrm{v}^{\frac{-\gamma}{\nu-\eta}-1}t^{\frac{\left(\eta-1 \right)\gamma}{\nu-\eta}+1}\right.\\
\left.-t^{0}\mathrm{v}^{\frac{-\gamma}{\nu-1}-1+\frac{1}{\nu-1}}\right), & \mathrm{v}>\mathrm{v}_c(t)\end{cases}
\end{equation}
Where critical velocity $\mathrm{v}_c(t)=\eta c\left(t\right)^{\nu-1}$.The scaling of $p(\mathrm{v},t)$ is $\eta$ dependent for this regime

\subsubsection{$\nu>1$}
For this parameter regime, we would again start with $\nu<\eta$

For fixed $t$, $p(\mathrm{v},t)$ would be  
\begin{equation}
p(\mathrm{v}, t)\approx \begin{cases}\frac{(\gamma-1)\tau_0^{(\gamma-1)} \left(\eta c\right)^{\frac{\gamma}{\nu-\eta}}}{(\eta-1)\gamma+(\nu-\eta)}\left(\mathrm{v}^{\frac{-\gamma}{\nu-\eta}-1}t^{\frac{\left(\eta-1 \right)\gamma}{\nu-\eta}+1}\right.\\
\left.-t^{0}\mathrm{v}^{\frac{-\gamma}{\nu-1}-1+\frac{1}{\nu-1}}\right) ,&\mathrm{v}<\mathrm{v}_c(t) \\ 0 , & \mathrm{v}>\mathrm{v}_c(t)\end{cases}
\end{equation}

Where critical velocity $\mathrm{v}_c(t)=\eta c\left(t\right)^{\nu-1}$
We would decrease the $\eta$ parameter till $\eta=\nu$
\begin{equation}
p(\mathrm{v}, t)\approx \begin{cases}\frac{(\gamma-1)\tau_0^{\gamma-1}}{\gamma(\eta-1)} t^{0}{\frac{\mathrm{v}}{\eta c}}^{-\frac{\gamma}{\eta-1}+\frac{1}{\eta-1}-1}, & \mathrm{v}<\mathrm{v}_c(t) \\ 0, &\mathrm{v}>\mathrm{v}_c(t)\end{cases}
\end{equation}
We would further decrease the $\eta$ parameter so that $\nu>\eta$
For fixed $t$, $p(\mathrm{v},t)$ would be  
\begin{equation}
p(\mathrm{v}, t)\approx \begin{cases}\frac{(\gamma-1)\tau_0^{(\gamma-1)} \left(\eta c\right)^{\frac{\gamma}{\nu-\eta}}}{(\eta-1)\gamma+(\nu-\eta)}\left(t^{0}\mathrm{v}^{\frac{-\gamma}{\nu-1}-1+\frac{1}{\nu-1}}\right), &\mathrm{v}<\mathrm{v}_c(t) \\ \frac{(\gamma-1)\tau_0^{(\gamma-1)} \left(\eta c\right)^{\frac{\gamma}{\nu-\eta}}}{(\eta-1)\gamma+(\nu-\eta)}\mathrm{v}^{\frac{-\gamma}{\nu-\eta}-1}\left(t^{1+\frac{\left(n-1 \right)\gamma}{\nu-\eta}}\right), &  \mathrm{v}>\mathrm{v}_c(t)\end{cases}
\end{equation}

Where critical velocity $\mathrm{v}_c(t)=\eta c\left(t\right)^{\nu-1}$.

The scaling of $p(\mathrm{v},t)$ is $\eta$ dependent for this regime

\section{Derivation of the Propagator}\label{Detailed_calculation_Propagator}
In this appendix, we present the detailed derivation of the velocity propagator
\( p(\mathrm{v},t) \) for Variable Speed Generalized Lévy Walks (VGLWs). Building on the renewal framework and joint time distributions introduced in
Appendix~A, we derive the full velocity propagator by integrating over all
possible elapsed times within a step. We obtain explicit asymptotic forms of
\( p(\mathrm{v},t) \).

\subsection{$\nu\neq\eta$}


 Eq.~\ref{eq:pvttprime} contains the Heaviside step function $\Theta(\cdot)$, which evaluates to unity only when $\left(\frac{\mathrm{v}}{\eta\, c\, t^{\prime\,\eta - 1}}\right)^{\frac{1}{\nu - \eta}} > t^{\prime}$
Depending on the specific choices of the parameters $\nu$ and $\eta$, four distinct cases arise, corresponding to different regions of parameter space. Each of these cases must be considered separately for both $\gamma < 1$ and $\gamma > 1$.  

\subsubsection{$\gamma<1$}

According to~\cite{akimoto2020infinite}, the Laplace transform of $R(t)$ is given by $
\widetilde{R}(s) = \frac{1}{1 - \widetilde{\psi}(s)}$,
where $s$ is the Laplace variable conjugate to $t$. The inverse Laplace transform  
$\mathcal{L}^{-1}\!\left[\frac{1}{1 - \widetilde{\psi}(s)}\right]$ is evaluated in  
Eq.~\ref{R(t)_g_l_1} for the case $\gamma < 1$.  
From that expression, the corresponding time-domain form of $R(t - t')$ is
\begin{eqnarray}\label{expansion_of_r(t)}
R(t-t^{\prime})&\approx&\frac{(t-t^\prime)^{\gamma-1}}{|\Gamma(1-\gamma)|\Gamma(\gamma)t_0^\gamma} \nonumber\\
&\approx& \frac{1}{|\Gamma(1-\gamma)|\Gamma(\gamma)t_0^\gamma}t^{\gamma-1}\left(1-\frac{t^{\prime}}{t}\right)^{\gamma-1}\nonumber\\
&\approx&\frac{1}{|\Gamma(1-\gamma)|\Gamma(\gamma)t_0^\gamma}\left(t^{\gamma-1}\left(1-\frac{(\gamma-1)t^{\prime}}{t}\right.\right.\nonumber\\
&&\left.\left.+\frac{(\gamma-1)(\gamma-2)t^{\prime2}}{2t^2}+O\left(\left(\frac{t^\prime}{t}\right)^2\right)\right)\right)\nonumber\\
\end{eqnarray}
This expansion of $R(t - t')$ will be used in the subsequent equations.
\paragraph{\textbf{Case 1: $\nu<1$ and $\nu>\eta$}}

Since $\nu > \eta$, the exponent $1/(\nu-\eta)$ is positive, and since 
$\nu - 1 < 0$, the function $t'^{\,\nu - 1}$ is monotonically decreasing. 
Thus the inequality$
\left(\frac{v}{\eta c\, t'^{\,\eta-1}}\right)^{\frac{1}{\nu-\eta}} > t'
$
is equivalently rewritten as

\begin{equation}\label{heaviside_1}
\Theta\!\left[\left(\frac{v}{\eta c\, t'^{\,\eta-1}}\right)^{\frac{1}{\nu-\eta}} - t'\right]
= 
\Theta\!\left(t' - \left(\frac{v}{\eta c}\right)^{\frac{1}{\nu-1}}\right).    
\end{equation}

\textbf{Sub-Case:($t<\left(\frac{\mathrm{v}}{\eta c}\right)^{\frac{1}{\nu-1}}$)}
In this limit, the argument of the Heaviside function (Eq.~\ref{heaviside_1}) is 0 for all $0 \le t' \le t$, so the integrand vanishes.

\textbf{Sub-Case:($t>\left(\frac{\mathrm{v}}{\eta c}\right)^{\frac{1}{\nu-1}}$)}
\begin{equation}
\begin{aligned}
    p\left(\mathrm{v},t\right) & =\int_{0}^t p\left(\mathrm{v},t, t^{\prime}\right) d t^{\prime}\\
 & \approx \frac{\gamma t_0^{\gamma}\left(\eta c\right)^{\frac{\gamma}{\nu-\eta}}}{(\nu-\eta)}\mathrm{v}^{\frac{-\gamma}{\nu-\eta}-1}\int_{0}^t R(t-t^{\prime})t^{\prime\frac{\left(\eta-1 \right)\gamma}{\nu-\eta}}\\
 & \times \Theta\left(t^{\prime}-\left(\frac{\mathrm{v}}{\eta c}\right)^{\frac{1}{\nu-1}}\right)dt^{\prime}\\  
 &\approx \frac{\gamma\left(\eta c\right)^{\frac{\gamma}{\nu-\eta}}}{|\Gamma(1-\gamma)|\Gamma(\gamma)(\nu-\eta)}\mathrm{v}^{\frac{-\gamma}{\nu-\eta}-1}\int_{\left(\frac{\mathrm{v}}{\eta c}\right)^{\frac{1}{\nu-1}}}^{t} (t-t^{\prime})^{\gamma-1}\\
 &\times t^{\prime\frac{\left(\eta-1 \right)\gamma}{\nu-\eta}} dt^{\prime}\\
\end{aligned}
\end{equation}
Using Eq.~\ref{expansion_of_r(t)} and simplifying, we obtain the following expression for $p(v,t)$. The contribution from the upper integration limit produces terms with identical powers of $t$, while from the lower limit we retain only the term with the highest power of $t$
\begin{equation}
\begin{aligned}
p(\mathrm{v},t)&\approx k_0 \mathrm{v}^{\frac{-\gamma}{\nu-\eta}-1}\left(t^{\gamma+\frac{\left(\eta-1 \right)\gamma}{\nu-\eta}}\right)\\
& -k_{1}t^{\gamma-1}\mathrm{v}^{\frac{-\gamma}{\nu-1}-1+\frac{1}{\nu-1}}\\
\end{aligned}
\end{equation}
 Where $k_0$ and $k_1$ are constants. 
\begin{equation}\label{value_of_constants}
\begin{aligned}
    k_0&= \frac{\gamma \left(\eta c\right)^{\frac{\gamma}{\nu-\eta}}}{|\Gamma(1-\gamma)|\Gamma(\gamma)((\eta-1)\gamma+(\nu-\eta))}\\
    &-\frac{(\gamma-1)\gamma\left(\eta c\right)^{\frac{\gamma}{\nu-\eta}}}{|\Gamma(1-\gamma)|\Gamma(\gamma)((\eta-1)\gamma+2(\nu-\eta))}+\ldots\\
    k_1&=\frac{\gamma\left(\eta c\right)^{\frac{\gamma}{\nu-1}-\frac{1}{\nu-1}}}{|\Gamma(1-\gamma)|\Gamma(\gamma)((\eta-1)\gamma+(\nu-\eta))}\\
\end{aligned}
\end{equation}

Summarising the above equation results, we can write $p(\mathrm{v},t)$ for ($\gamma<1,\nu<1\&\nu>\eta$)

\begin{equation}
p(\mathrm{v}, t)\approx \begin{cases}0, & t<t_c(\mathrm{v}) \\ k_0 \mathrm{v}^{\frac{-\gamma}{\nu-\eta}-1}\left(t^{\gamma+\frac{\left(n-1 \right)\gamma}{\nu-\eta}}\right)\\-k_{1}t^{\gamma-1}\mathrm{v}^{\frac{-\gamma}{\nu-1}-1+\frac{1}{\nu-1}}, & t>t_c(\mathrm{v})\end{cases}
\end{equation}

Where critical time $t_c(\mathrm{v})=\left(\frac{\mathrm{v}}{\eta c}\right)^{\frac{1}{\nu-1}}$

For fixed $t$, $p(\mathrm{v},t)$ would be  
\begin{equation}
p(\mathrm{v}, t)\approx \begin{cases}0, & \mathrm{v}<\mathrm{v}_c(t) \\ k_0 \mathrm{v}^{\frac{-\gamma}{\nu-\eta}-1}\left(t^{\gamma+\frac{\left(n-1 \right)\gamma}{\nu-\eta}}\right)\\-k_{1}t^{\gamma-1}\mathrm{v}^{\frac{-\gamma}{\nu-1}-1+\frac{1}{\nu-1}}, & \mathrm{v}>\mathrm{v}_c(t)\end{cases}
\end{equation}

Where critical velocity $\mathrm{v}_c(t)=\eta c\left(t\right)^{\nu-1}$

\paragraph{\textbf{Case 2: $\nu<1$ and $\nu<\eta$}}
Since in this case $\nu < 1$ and $\nu < \eta$, the inequality 
$\left(\frac{v}{\eta c\, t'^{\,\eta-1}}\right)^{\!\frac{1}{\nu-\eta}} > t'$ 
is equivalent to 
$t' < \left(\frac{v}{\eta c}\right)^{\!\frac{1}{\nu-1}}$.
Therefore, the Heaviside function of Eq.~\ref{eq:pvttprime} can be rewritten as 
\begin{equation}\label{heaviside_theta_2}
\Theta\!\left[\left(\frac{v}{\eta c\, t'^{\,\eta-1}}\right)^{\frac{1}{\nu-\eta}} - t'\right]
= \Theta\!\left[\left(\frac{v}{\eta c}\right)^{\frac{1}{\nu-1}} - t'\right]
\end{equation}

\textbf{Sub-Case:($t<\left(\frac{\mathrm{v}}{\eta c}\right)^{\frac{1}{\nu-1}}$)}
\begin{equation}
\begin{aligned}
p\left(\mathrm{v},t\right) & =\int_{0}^t p\left(\mathrm{v},t, t^{\prime}\right) d t^{\prime}\\
& \approx \frac{\gamma t_0^{\gamma}\left(\eta c\right)^{\frac{\gamma}{\nu-\eta}}}{(\nu-\eta)}\mathrm{v}^{\frac{-\gamma}{\nu-\eta}-1}\int_{0}^t R(t-t^{\prime})t^{\prime\frac{\left(n-1 \right)\gamma}{\nu-\eta}}\\
&\times\Theta\left(\left(\frac{\mathrm{v}}{\eta c}\right)^{\frac{1}{\nu-1}}-t^{\prime}\right)dt^{\prime}\\
& \approx \frac{\gamma\left(\eta c\right)^{\frac{\gamma}{\nu-\eta}}}{|\Gamma(1-\gamma)|\Gamma(\gamma)(\nu-\eta)}\mathrm{v}^{\frac{-\gamma}{\nu-\eta}-1}\int_{0}^{t}\left(t-t^\prime\right)^{\gamma-1}\\
&\times t^{\prime\frac{\left(n-1 \right)\gamma}{\nu-\eta}} dt^{\prime}\\
&\approx k_0\mathrm{v}^{\frac{-\gamma}{\nu-\eta}-1}\left(t^{\gamma+\frac{\left(n-1 \right)\gamma}{\nu-\eta}}\right)
\end{aligned}
\end{equation}
Where $k_0$ is a constant and the value of $k_0$ is described in Eq.~\ref{value_of_constants}. 

\textbf{Sub-Case:($t>\left(\frac{\mathrm{v}}{\eta c}\right)^{\frac{1}{\nu-1}}$)}
\begin{equation}
\begin{aligned}
  p\left(\mathrm{v},t\right) & =\int_{0}^t p\left(\mathrm{v},t, t^{\prime}\right) d t^{\prime}\\
 & \approx \frac{\gamma t_0^{\gamma}\left(\eta c\right)^{\frac{\gamma}{\nu-\eta}}}{(\nu-\eta)}\mathrm{v}^{\frac{-\gamma}{\nu-\eta}-1}\int_{0}^t R(t-t^{\prime})t^{\prime\frac{\left(\eta-1 \right)\gamma}{\nu-\eta}}\\
 & \times \Theta\left(t^{\prime}-\left(\frac{\mathrm{v}}{\eta c}\right)^{\frac{1}{\nu-1}}\right)dt^{\prime}\\  
 &\approx \frac{\gamma\left(\eta c\right)^{\frac{\gamma}{\nu-\eta}}}{|\Gamma(1-\gamma)|\Gamma(\gamma)(\nu-\eta)}\mathrm{v}^{\frac{-\gamma}{\nu-\eta}-1}\int_{0}^{\left(\frac{\mathrm{v}}{\eta c}\right)^{\frac{1}{\nu-1}}} (t-t^{\prime})^{\gamma-1}\\
 &\times t^{\prime\frac{\left(\eta-1 \right)\gamma}{\nu-\eta}} dt^{\prime}\\
\end{aligned}
\end{equation}
Simplifying and only retaining the highest power of $t$, we would get 
\begin{equation}
\begin{aligned}
p(\mathrm{v},t)&\approx k_{1}t^{\gamma-1}\mathrm{v}^{\frac{-\gamma}{\nu-1}-1+\frac{1}{\nu-1}}
\end{aligned}
\end{equation}
Where $k_1$ is a constant and given by Eq.~\ref{value_of_constants}. Summarising the above equation results, we can write $p(\mathrm{v},t)$ for ($\gamma<1,\nu<1\&\nu>\eta$)

\begin{equation}
p(\mathrm{v}, t)\approx \begin{cases}k_0\mathrm{v}^{\frac{-\gamma}{\nu-\eta}-1}\left(t^{\gamma+\frac{\left(n-1 \right)\gamma}{\nu-\eta}}\right), &  t<t_c(\mathrm{v}) \\k_{1}t^{\gamma-1}\mathrm{v}^{\frac{-\gamma}{\nu-1}-1
+\frac{1}{\nu-1}},
 & t>t_c(\mathrm{v})\end{cases}
\end{equation}

Where critical time $t_c(\mathrm{v})=\left(\frac{\mathrm{v}}{\eta c}\right)^{\frac{1}{\nu-1}}$

For fixed $t$, $p(\mathrm{v},t)$ would be  
\begin{equation}
p(\mathrm{v}, t)\approx \begin{cases}k_0\mathrm{v}^{\frac{-\gamma}{\nu-\eta}-1}\left(t^{\gamma+\frac{\left(n-1 \right)\gamma}{\nu-\eta}}\right), & \mathrm{v}<\mathrm{v}_c(t) \\ k_{1}t^{\gamma-1}\mathrm{v}^{\frac{-\gamma}{\nu-1}-1+\frac{1}{\nu-1}},&\mathrm{v}>\mathrm{v}_c(t)\end{cases}
\end{equation}

Where critical velocity $\mathrm{v}_c(t)=\eta c\left(t\right)^{\nu-1}$

\paragraph{\textbf{Case 3: $\nu>1$ and $\nu>\eta$}}
Since $\nu > 1$ and $\nu > \eta$, the inequality 
$\left(\frac{v}{\eta c\, t'^{\,\eta-1}}\right)^{\!\frac{1}{\nu-\eta}} > t'$ 
reduces to 
$t' < \left(\frac{v}{\eta c}\right)^{\!\frac{1}{\nu-1}}$.
Hence,$
\Theta\!\left[\left(\frac{v}{\eta c\, t'^{\eta-1}}\right)^{\frac{1}{\nu-\eta}} - t'\right]
= \Theta\!\left[\left(\frac{v}{\eta c}\right)^{\frac{1}{\nu-1}} - t'\right].
$
For this case, the calculation proceeds analogously to Case~2 
($\nu < 1$ and $\eta > \nu$) when $\mathrm{v}$ is held fixed. 
However, when $t$ is fixed, the functional form of $p(\mathrm{v},t)$is interchanged because here $\nu > 1$.

$p(\mathrm{v},t)$ for fixed $\mathrm{v}$ ($\gamma<1,\nu>1$ \& $\nu>\eta$)

\begin{equation}
p(\mathrm{v}, t)\approx \begin{cases}k_0\mathrm{v}^{\frac{-\gamma}{\nu-\eta}-1}\left(t^{\gamma+\frac{\left(n-1 \right)\gamma}{\nu-\eta}}\right), & t<t_c(\mathrm{v}) \\k_{1}t^{\gamma-1}\mathrm{v}^{\frac{-\gamma}{\nu-1}-1
+\frac{1}{\nu-1}},
 & t>t_c(\mathrm{v})\end{cases}
\end{equation}

Where critical time $t_c(\mathrm{v})=\left(\frac{\mathrm{v}}{\eta c}\right)^{\frac{1}{\nu-1}}$ and $k_0,k_1$ are constants and described by Eq.~\ref{value_of_constants}.

For fixed $t$, the functional form of $p(v,t)$ is 
\begin{equation}
p(\mathrm{v}, t)\approx \begin{cases}
k_{1}t^{\gamma-1}\mathrm{v}^{\frac{-\gamma}{\nu-1}-1+\frac{1}{\nu-1}},& \mathrm{v}<\mathrm{v}_c(t) \\k_0\mathrm{v}^{\frac{-\gamma}{\nu-\eta}-1}\left(t^{\gamma+\frac{\left(n-1 \right)\gamma}{\nu-\eta}}\right), &\mathrm{v}>\mathrm{v}_c(t)\end{cases}
\end{equation}

Where critical velocity $\mathrm{v}_c(t)=\eta c\left(t\right)^{\nu-1}$

\paragraph{\textbf{Case 4: $\nu>1$ and $\nu<\eta$}}
Since $\nu > 1$ and $\nu > \eta$, the inequality 
$\left(\frac{v}{\eta c\, t'^{\,\eta-1}}\right)^{\!\frac{1}{\nu-\eta}} > t'$ 
reduces to 
$t' > \left(\frac{v}{\eta c}\right)^{\!\frac{1}{\nu-1}}$.
Thus,$
\Theta\!\left[\left(\frac{v}{\eta c\, t'^{\eta-1}}\right)^{\frac{1}{\nu-\eta}} - t'\right]
= \Theta\!\left[t' - \left(\frac{v}{\eta c}\right)^{\frac{1}{\nu-1}}\right]$
For this case, the calculation is analogous to Case~1 ($\nu < 1$ and $\eta < \nu$) when $v$ is held fixed. 
However, for fixed $t$, the functional form is reversed because $\nu > 1$.

$p(\mathrm{v},t)$ for ($\gamma<1,\nu>1\&\nu<\eta$)

\begin{equation}
p(\mathrm{v}, t)\approx \begin{cases}0, & t<t_c(\mathrm{v}) \\ k_0 \mathrm{v}^{\frac{-\gamma}{\nu-\eta}-1}\left(t^{\gamma+\frac{\left(n-1 \right)\gamma}{\nu-\eta}}\right)\\-k_{1}t^{\gamma-1}\mathrm{v}^{\frac{-\gamma}{\nu-1}-1+\frac{1}{\nu-1}}, &  t>t_c(\mathrm{v})\end{cases}
\end{equation}

Where critical time $t_c(\mathrm{v})=\left(\frac{\mathrm{v}}{\eta c}\right)^{\frac{1}{\nu-1}}$ and $k_0,k_1$ are constants and given by Eq.~\ref{value_of_constants}.

For fixed $t$, $p(\mathrm{v},t)$ would be  
\begin{equation}
p(\mathrm{v}, t)\approx \begin{cases}k_0 \mathrm{v}^{\frac{-\gamma}{\nu-\eta}-1}\left(t^{\gamma+\frac{\left(n-1 \right)\gamma}{\nu-\eta}}\right)\\-k_{1}t^{\gamma-1}\mathrm{v}^{\frac{-\gamma}{\nu-1}-1+\frac{1}{\nu-1}}, & \mathrm{v}<\mathrm{v}_c(t) \\ 0,  & \mathrm{v}>\mathrm{v}_c(t)\end{cases}
\end{equation}
Where critical velocity $\mathrm{v}_c(t)=\eta c\left(t\right)^{\nu-1}$.
\subsubsection{$\gamma>1$}
For $\gamma > 1$, the renewal function reduces to 
$R(t - t') \approx \frac{(\gamma - 1)(t - t')^{0}}{\gamma t_0}$,
as given in Eq.~\ref{R(t)_g_g_1}. Apart from this change, the calculation proceeds in the same manner as in the 
$\gamma < 1$ case.

\paragraph{\textbf{Case 1: $\nu<1$ and $\nu>\eta$}}
For this parameter choices, the Heaviside Theta function in Eq.~\ref{eq:pvttprime} would be similar as Eq.~\ref{heaviside_1}.

\textbf{Sub-Case:($t<\left(\frac{\mathrm{v}}{\eta c}\right)^{\frac{1}{\nu-1}}$)}
In this limit, the argument of the Heaviside function (Eq.~\ref{heaviside_1}) is 0 for all $0 \le t' \le t$, so the integrand vanishes.

\textbf{Sub-Case 2:($t>\left(\frac{\mathrm{v}}{\eta c}\right)^{\frac{1}{\nu-1}}$)}
\begin{equation}
\begin{aligned}
    p\left(\mathrm{v},t\right) & =\int_{0}^t p\left(\mathrm{v},t, t^{\prime}\right) d t^{\prime}\\
& \approx \frac{\gamma\tau_0^{\gamma}\left(\eta c\right)^{\frac{\gamma}{\nu-\eta}}}{\nu-\eta}\mathrm{v}^{\frac{-\gamma}{\nu-\eta}-1}\int_{0}^t R(t-t^{\prime})t^{\prime\frac{\left(\eta-1 \right)\gamma}{\nu-\eta}}\\
&\times\Theta\left(t^{\prime}-\left(\frac{\mathrm{v}}{\eta c}\right)^{\frac{1}{\nu-1}}\right)dt^{\prime}\\
& \approx \frac{(\gamma-1)\tau_0^{(\gamma-1)}\left(\eta c\right)^{\frac{\gamma}{\nu-\eta}}}{\nu-\eta}\mathrm{v}^{\frac{-\gamma}{\nu-\eta}-1}\int_{\left(\frac{\mathrm{v}}{\eta c}\right)^{\frac{1}{\nu-1}}}^{t} t^0\\
&\times t^{\prime\frac{\left(\eta-1 \right)\gamma}{\nu-\eta}} dt^{\prime}\\
 \end{aligned}   
\end{equation}
Simplifying, we would get 
\begin{equation}
\begin{aligned}
p(\mathrm{v},t)&\approx\frac{(\gamma-1)\tau_0^{(\gamma-1)} \left(\eta c\right)^{\frac{\gamma}{\nu-\eta}}}{(\eta-1)\gamma+(\nu-\eta)}\left(\mathrm{v}^{\frac{-\gamma}{\nu-\eta}-1}t^{\frac{\left(\eta-1 \right)\gamma}{\nu-\eta}+1}\right.\\
&\left.-t^{0}\mathrm{v}^{\frac{-\gamma}{\nu-1}-1+\frac{1}{\nu-1}}\right)
\end{aligned}
\end{equation}

Summarising the above equation results, we can write $p(\mathrm{v},t)$ for ($\gamma>1,\nu<1\&\nu>\eta$)

\begin{equation}
p(\mathrm{v}, t)\approx \begin{cases}0, & t<t_c(\mathrm{v}) \\ \frac{(\gamma-1)\tau_0^{(\gamma-1)} \left(\eta c\right)^{\frac{\gamma}{\nu-\eta}}}{(\eta-1)\gamma+(\nu-\eta)}\left(\mathrm{v}^{\frac{-\gamma}{\nu-\eta}-1}t^{\frac{\left(\eta-1 \right)\gamma}{\nu-\eta}+1}\right.\\
\left.-t^{0}\mathrm{v}^{\frac{-\gamma}{\nu-1}-1+\frac{1}{\nu-1}}\right) ,&t>t_c(\mathrm{v})\end{cases}
\end{equation}
Where critical time $t_c(\mathrm{v})=\left(\frac{\mathrm{v}}{\eta c}\right)^{\frac{1}{\nu-1}}$

For fixed $t$, $p(\mathrm{v},t)$ would be  
\begin{equation}
p(\mathrm{v}, t)\approx \begin{cases}0, &\mathrm{v}<\mathrm{v}_c(t) \\ \frac{(\gamma-1)\tau_0^{(\gamma-1)} \left(\eta c\right)^{\frac{\gamma}{\nu-\eta}}}{(\eta-1)\gamma+(\nu-\eta)}\left(\mathrm{v}^{\frac{-\gamma}{\nu-\eta}-1}t^{\frac{\left(\eta-1 \right)\gamma}{\nu-\eta}+1}\right.\\
\left.-t^{0}\mathrm{v}^{\frac{-\gamma}{\nu-1}-1+\frac{1}{\nu-1}}\right), &  \mathrm{v}>\mathrm{v}_c(t)\end{cases}
\end{equation}
\paragraph{\textbf{Case 2: $\nu<1$ and $\nu<\eta$\\}}
For this parameter choices, the Heaviside function of Eq.~\ref{eq:pvttprime} can be rewritten as Eq.~\ref{heaviside_theta_2}.

\textbf{Sub-Case 1:($t<\left(\frac{\mathrm{v}}{\eta c}\right)^{\frac{1}{\nu-1}}$)}
\begin{equation}
\begin{aligned}
    p\left(\mathrm{v},t\right) & =\int_{0}^t p\left(\mathrm{v},t, t^{\prime}\right) d t^{\prime}\\
& \approx \frac{\gamma\tau_0^{\gamma} \left(\eta c\right)^{\frac{\gamma}{\nu-\eta}}}{\nu-\eta}\mathrm{v}^{\frac{-\gamma}{\nu-\eta}-1}\int_{0}^t R(t-t^{\prime})t^{\prime\frac{\left(n-1 \right)\gamma}{\nu-\eta}}\\
&\times\Theta\left(\left(\frac{\mathrm{v}}{\eta c}\right)^{\frac{1}{\nu-1}}-t^{\prime}\right)dt^{\prime}\\
& \approx \frac{(\gamma-1) \tau_0^{\gamma-1} \left(\eta c\right)^{\frac{\gamma}{\nu-\eta}}}{\nu-\eta}\mathrm{v}^{\frac{-\gamma}{\nu-\eta}-1}t^0\int_{0}^{t} t^{\prime\frac{\left(n-1 \right)\gamma}{\nu-\eta}} dt^{\prime}\\
\end{aligned}
\end{equation}
\textbf{Sub-Case 2:($t>\left(\frac{\mathrm{v}}{\eta c}\right)^{\frac{1}{\nu-1}}$)}
\begin{equation}
\begin{aligned}
 p\left(\mathrm{v},t\right) & =\int_{0}^t p\left(\mathrm{v},t, t^{\prime}\right) d t^{\prime}\\
& \approx \frac{\gamma \tau_0^{\gamma} \left(\eta c\right)^{\frac{\gamma}{\nu-\eta}}}{\nu-\eta}\mathrm{v}^{\frac{-\gamma}{\nu-\eta}-1}\int_{0}^t R(t-t^{\prime})t^{\prime\frac{\left(n-1 \right)\gamma}{\nu-\eta}}\\
&\times\Theta\left(\left(\frac{\mathrm{v}}{\eta c}\right)^{\frac{1}{\nu-1}}-t^{\prime}\right)dt^{\prime}\\
& \approx \frac{(\gamma-1) \tau_0^{\gamma-1} \left(\eta c\right)^{\frac{\gamma}{\nu-\eta}}}{\nu-\eta}\mathrm{v}^{\frac{-\gamma}{\nu-\eta}-1}t^0\int_{0}^{\left(\frac{\mathrm{v}}{\eta c}\right)^{\frac{1}{\nu-1}}}\\
&\times t^{\prime\frac{\left(n-1 \right)\gamma}{\nu-\eta}} dt^{\prime}\\
\end{aligned}
\end{equation}
Summarising the above equation results, we can write $p(\mathrm{v},t)$ for ($\gamma>1,\nu<1\&\nu>\eta$)

\begin{equation}
p(\mathrm{v}, t)\approx \begin{cases}\frac{(\gamma-1)\tau_0^{(\gamma-1)} \left(\eta c\right)^{\frac{\gamma-1}{\nu-1}}}{(\eta-1)\gamma+(\nu-\eta)}\mathrm{v}^{\frac{-\gamma}{\nu-\eta}-1}\left(t^{1+\frac{\left(n-1 \right)\gamma}{\nu-\eta}}\right), & t<t_c(\mathrm{v}) \\ \frac{(\gamma-1)\tau_0^{(\gamma-1)} \left(\eta c\right)^{\frac{\gamma}{\nu-\eta}}}{(\eta-1)\gamma+(\nu-\eta)}\left(t^{0}\mathrm{v}^{\frac{-\gamma}{\nu-1}-1+\frac{1}{\nu-1}}\right), & t>t_c(\mathrm{v})\end{cases}
\end{equation}
Where critical time $t_c(\mathrm{v})=\left(\frac{\mathrm{v}}{\eta c}\right)^{\frac{1}{\nu-1}}$

For fixed $t$, $p(\mathrm{v},t)$ would be  
\begin{equation}
p(\mathrm{v}, t)\approx \begin{cases}\frac{(\gamma-1)\tau_0^{(\gamma-1)} \left(\eta c\right)^{\frac{\gamma}{\nu-\eta}}}{(\eta-1)\gamma+(\nu-\eta)}\mathrm{v}^{\frac{-\gamma}{\nu-\eta}-1}\left(t^{1+\frac{\left(n-1 \right)\gamma}{\nu-\eta}}\right), & \mathrm{v}<\mathrm{v}_c(t) \\ \frac{(\gamma-1)\tau_0^{(\gamma-1)} \left(\eta c\right)^{\frac{\gamma}{\nu-\eta}}}{(\eta-1)\gamma+(\nu-\eta)}\left(t^{0}\mathrm{v}^{\frac{-\gamma}{\nu-1}-1+\frac{1}{\nu-1}}\right), & \mathrm{v}>\mathrm{v}_c(t)\end{cases}
\end{equation}
Where critical velocity $\mathrm{v}_c(t)=\eta c\left(t\right)^{\nu-1}$
\paragraph{\textbf{Case 3: $\nu>1$ and $\nu>\eta$}}
For this case, the calculation would be similar to Case 2: $\nu<1$ and $\eta>\nu$ for fixed $\mathrm{v}$. Although, the functional form would be interchanged for fixed $t$ as $\nu>1$.

$p(\mathrm{v},t)$ for ($\gamma<1,\nu<1\&\nu>\eta$)

\begin{equation}
p(\mathrm{v}, t)\approx \begin{cases}\frac{(\gamma-1)\tau_0^{(\gamma-1)} \left(\eta c\right)^{\frac{\gamma-1}{\nu-1}}}{(\eta-1)\gamma+(\nu-\eta)}\mathrm{v}^{\frac{-\gamma}{\nu-\eta}-1}\left(t^{1+\frac{\left(n-1 \right)\gamma}{\nu-\eta}}\right), & t<t_c(\mathrm{v}) \\ \frac{(\gamma-1)\tau_0^{(\gamma-1)} \left(\eta c\right)^{\frac{\gamma}{\nu-\eta}}}{(\eta-1)\gamma+(\nu-\eta)}\left(t^{0}\mathrm{v}^{\frac{-\gamma}{\nu-1}-1+\frac{1}{\nu-1}}\right), & t>t_c(\mathrm{v})\end{cases}
\end{equation}
Where critical time $t_c(\mathrm{v})=\left(\frac{\mathrm{v}}{\eta c}\right)^{\frac{1}{\nu-1}}$

For fixed $t$, $p(\mathrm{v},t)$ would be  
\begin{equation}
p(\mathrm{v}, t)\approx \begin{cases}\frac{(\gamma-1)\tau_0^{(\gamma-1)} \left(\eta c\right)^{\frac{\gamma}{\nu-\eta}}}{(\eta-1)\gamma+(\nu-\eta)}\left(t^{0}\mathrm{v}^{\frac{-\gamma}{\nu-1}-1+\frac{1}{\nu-1}}\right), & \mathrm{v}<\mathrm{v}_c(t) \\ \frac{(\gamma-1)\tau_0^{(\gamma-1)} \left(\eta c\right)^{\frac{\gamma}{\nu-\eta}}}{(\eta-1)\gamma+(\nu-\eta)}\mathrm{v}^{\frac{-\gamma}{\nu-\eta}-1}\left(t^{1+\frac{\left(n-1 \right)\gamma}{\nu-\eta}}\right), & \mathrm{v}>\mathrm{v}_c(t)\end{cases}
\end{equation}

Where critical velocity $\mathrm{v}_c(t)=\eta c\left(t\right)^{\nu-1}$
\paragraph{\textbf{Case 4: $\nu>1$ and $\nu<\eta$}}
For this case, the calculation would be similar to Case 1: $\nu<1$ and $\eta<\nu$ for fixed $\mathrm{v}$. Although, the functional form would be interchanged for fixed $t$ as $\nu>1$

$p(\mathrm{v},t)$ for ($\gamma<1,\nu>1\&\nu<\eta$)

\begin{equation}
p(\mathrm{v}, t)\approx \begin{cases}0, & \mathrm{v}<\mathrm{v}_c(t) \\ \frac{(\gamma-1)\tau_0^{(\gamma-1)} \left(\eta c\right)^{\frac{\gamma}{\nu-\eta}}}{(\eta-1)\gamma+(\nu-\eta)}\left(\mathrm{v}^{\frac{-\gamma}{\nu-\eta}-1}t^{\frac{\left(\eta-1 \right)\gamma}{\nu-\eta}+1}\right.\\
\left.-t^{0}\mathrm{v}^{\frac{-\gamma}{\nu-1}-1+\frac{1}{\nu-1}}\right), & \mathrm{v}>\mathrm{v}_c(t)\end{cases}
\end{equation}

Where critical time $t_c(\mathrm{v})=\left(\frac{\mathrm{v}}{\eta c}\right)^{\frac{1}{\nu-1}}$ and $k_0,k_1,k_2,k_3$ are constants.

For fixed $t$, $p(\mathrm{v},t)$ would be  
\begin{equation}
p(\mathrm{v}, t)\approx \begin{cases}\frac{(\gamma-1)\tau_0^{(\gamma-1)} \left(\eta c\right)^{\frac{\gamma}{\nu-\eta}}}{(\eta-1)\gamma+(\nu-\eta)}\left(\mathrm{v}^{\frac{-\gamma}{\nu-\eta}-1}t^{\frac{\left(\eta-1 \right)\gamma}{\nu-\eta}+1}\right.\\
\left.-t^{0}\mathrm{v}^{\frac{-\gamma}{\nu-1}-1+\frac{1}{\nu-1}}\right), & \mathrm{v}<\mathrm{v}_c(t) \\ 0,  & \mathrm{v}>\mathrm{v}_c(t)\end{cases}
\end{equation}

Where critical velocity $\mathrm{v}_c(t)=\eta c\left(t\right)^{\nu-1}$
\subsection{Special case $\nu=\eta$}
For this case, the velocity is dependent only on $t^\prime$ or backward recurrence time
\begin{equation}
\mathrm{v}_{\nu,\eta}=\pm\eta c {t^\prime}^{\eta-1}  
\end{equation}
As the velocity is not explicitly dependent of $\tau$, the velocity at $t^\prime$ would be same for all steps greater than $t^\prime$. Based on this, we can write the propagator $p(\mathrm{v},t,t^\prime)$ for $\nu=\eta$ 
\begin{equation}
\begin{aligned}
p(\mathrm{v},t,t^\prime)&=R(t-t^\prime) \delta(|\mathrm{v}|-\eta c {t^\prime}^{\eta-1})\theta(t-t^\prime)\int_{t^\prime}^\infty \psi(t)dt
\end{aligned}    
\end{equation}

where $\int_{t^\prime}^\infty \psi(t)dt$ is the probability to get a step where the flight time is higher than $t^\prime$

To get $p(\mathrm{v},t)$, we have to integrate $p(\mathrm{v},t,t^\prime)$ thorough all $t^\prime$

For $\gamma<1$, 
\begin{equation}
\begin{aligned}
    p(\mathrm{v},t,t^\prime)&\approx \frac{1}{\gamma t_0^\gamma \Gamma|1-\gamma|\Gamma|\gamma|}(t-t^\prime)^{\gamma-1} \delta(|\mathrm{v}|-\eta c {t^\prime}^{\eta-1})\theta(t-t^\prime)\\
    &\times\int_{t^\prime}^\infty \psi(t)dt\\
    &\approx \frac{1}{\Gamma|1-\gamma|\Gamma|\gamma|} t^{\gamma-1}{t^\prime}^{-\gamma}\delta(|\mathrm{v}|-\eta c {t^\prime}^{\eta-1})\theta(t-t^\prime)
\end{aligned}    
\end{equation}
\begin{equation}
\begin{aligned}
    p(\mathrm{v},t)&=\int_0^\infty p(\mathrm{v},t,t^\prime) dt^\prime\\
    &\approx \frac{1}{(\eta-1)\gamma \Gamma|1-\gamma|\Gamma|\gamma|} t^{\gamma-1}{\frac{\mathrm{v}}{\eta c}}^{-\frac{\gamma}{\eta-1}+\frac{1}{\eta-1}-1}\\
    &\times\theta\left(t-\left(\frac{|\mathrm{v}|}{\eta c}\right)^\frac{1}{\eta-1}\right)
\end{aligned}    
\end{equation}
Summarising the above equation result, we can write $p(\mathrm{v},t)$ 

\begin{equation}
p(\mathrm{v}, t)\approx \begin{cases}0, & t<t_c(\mathrm{v}) \\ \frac{1}{(\eta-1)\gamma \Gamma|1-\gamma|\Gamma|\gamma|} t^{\gamma-1}{\frac{\mathrm{v}}{\eta c}}^{-\frac{\gamma}{\eta-1}+\frac{1}{\eta-1}-1}, & t>t_c(\mathrm{v})\end{cases}
\end{equation}
where critical time $t_c(\mathrm{v})=\left(\frac{\mathrm{v}}{\eta c}\right)^{\frac{1}{\eta-1}}$

For fixed $t$ and for $\eta<1$, $p(\mathrm{v},t)$ would be  

\begin{equation}
p(\mathrm{v}, t)\approx \begin{cases} 0,  & \mathrm{v}<\mathrm{v}_c(t) \\ \frac{1}{(\eta-1)\gamma \Gamma|1-\gamma|\Gamma|\gamma|} t^{\gamma-1}{\frac{\mathrm{v}}{\eta c}}^{-\frac{\gamma}{\eta-1}+\frac{1}{\eta-1}-1}, &  \mathrm{v}>\mathrm{v}_c(t)\end{cases}
\end{equation}
for $\eta>1$, $p(\mathrm{v},t)$ would be  
\begin{equation}
p(\mathrm{v}, t)\approx \begin{cases}\frac{1}{(\eta-1)\gamma \Gamma|1-\gamma|\Gamma|\gamma|} t^{\gamma-1}{\frac{\mathrm{v}}{\eta c}}^{-\frac{\gamma}{\eta-1}+\frac{1}{\eta-1}-1}, & \mathrm{v}<\mathrm{v}_c(t) \\ 0, & \mathrm{v}>\mathrm{v}_c(t)\end{cases}
\end{equation}

where critical velocity $\mathrm{v}_c(t)=\eta c\left(t\right)^{\eta-1}$

For $\gamma>1$, 
\begin{equation}
\begin{aligned}
    p(\mathrm{v},t,t^\prime)&\approx\frac{\gamma-1}{\gamma t_0}(t-t^\prime)^{0} \delta(|\mathrm{v}|-\eta c {t^\prime}^{\eta-1})\theta(t-t^\prime)\int_{t^\prime}^\infty \psi(t)dt\\
    &\approx\frac{(\gamma-1)\tau_0^{\gamma-1}}{\gamma} t^{0}{t^\prime}^{-\gamma}\delta(|\mathrm{v}|-\eta c {t^\prime}^{\eta-1})\theta(t-t^\prime)
\end{aligned}    
\end{equation}
\begin{equation}
\begin{aligned}
    p(\mathrm{v},t)&=\int_0^\infty p(\mathrm{v},t,t^\prime) dt^\prime\\
    &\approx\frac{(\gamma-1)\tau_0^{\gamma-1}}{\gamma(\eta-1)} t^{0}{\frac{\mathrm{v}}{\eta c}}^{-\frac{\gamma}{\eta-1}+\frac{1}{\eta-1}-1}\theta\left(t-\left(\frac{|\mathrm{v}|}{\eta c}\right)^\frac{1}{\eta-1}\right)
\end{aligned}    
\end{equation}
Summarising the above equation result, we can write $p(\mathrm{v},t)$ 
\begin{equation}
p(\mathrm{v}, t)\approx \begin{cases}0, & t<t_c(\mathrm{v}) \\ \frac{(\gamma-1)\tau_0^{\gamma-1}}{\gamma(\eta-1)} t^{0}{\frac{\mathrm{v}}{\eta c}}^{-\frac{\gamma}{\eta-1}+\frac{1}{\eta-1}-1}, & t>t_c(\mathrm{v})\end{cases}
\end{equation}
where critical time $t_c(\mathrm{v})=\left(\frac{\mathrm{v}}{\eta c}\right)^{\frac{1}{\eta-1}}$.
For fixed $t$ and for $\eta<1$, $p(\mathrm{v},t)$ would be  

\begin{equation}
p(\mathrm{v}, t)\approx \begin{cases} 0,  & \mathrm{v}<\mathrm{v}_c(t) \\ \frac{(\gamma-1)\tau_0^{\gamma-1}}{\gamma(\eta-1)} t^{0}{\frac{\mathrm{v}}{\eta c}}^{-\frac{\gamma}{\eta-1}+\frac{1}{\eta-1}-1}, & \mathrm{v}>\mathrm{v}_c(t)\end{cases}
\end{equation}
for $\eta>1$, $p(\mathrm{v},t)$ would be  
\begin{equation}
p(\mathrm{v}, t)\approx \begin{cases}\frac{(\gamma-1)\tau_0^{\gamma-1}}{\gamma(\eta-1)} t^{0}{\frac{\mathrm{v}}{\eta c}}^{-\frac{\gamma}{\eta-1}+\frac{1}{\eta-1}-1}, & \mathrm{v}<\mathrm{v}_c(t) \\ 0, &\mathrm{v}>\mathrm{v}_c(t)\end{cases}
\end{equation}
where critical velocity $\mathrm{v}_c(t)=\eta c\left(t\right)^{\eta-1}$

\begin{widetext}
\section{Derivation of the velocity correlation function}\label{vcf_calculation}
VCF $\langle \mathrm{v}(t)\mathrm{v}(t+\Delta)\rangle$ can be written as,
\begin{equation}
\langle \mathrm{v}(t)\mathrm{v}(t+\Delta)\rangle = C(t, \Delta) = \sum_{n=0}^{\infty} C_n(t, \Delta).
\end{equation}
where $C_n(t, \Delta) = \langle \mathrm{v}(t)\mathrm{v}(t+\Delta)\rangle_n$ the contribution to the VCF from trajectories that undergo exactly $n$ renewal events in the interval $(0, t)$. $C_n(s, u)$ can be expressed as 
\begin{equation}
C_n(s, u)=\int_0^{\infty} e^{-s t} d t \int_0^{\infty} e^{-u \Delta} d \Delta\langle \mathrm{v}(t) \mathrm{v}(t+\Delta)\rangle_n    
\end{equation}
Putting the constraint of $t_n<t<t+\Delta<t_{n+1}$ in the above equation, where $t_n$ and $t_{n+1}$ are the time of $n$ and $n+1$th renewal process, we get

\begin{eqnarray}
C_n(s, u)&=&\eta^2c^2\left\langle\tau_n^{2(\nu-\eta)}\int_0^{\infty} e^{-s t} d t \int_0^{\infty} e^{-u \Delta} d \Delta \left(t-t_n\right)^{\eta-1}\left(t+\Delta-t_n\right)^{\eta-1}\right.\theta\left(t_n<t<t_{n+1}\right)\nonumber\\ 
&&\left. \theta\left(t_n<t+\Delta<t_{n+1}\right)\right\rangle \end{eqnarray}
Here $\theta(\cdots)=1$ if the condition in the parenthesis is valid otherwise the $\theta$ function is zero, so $\theta(\ldots)$ is a square pulse function. The average is with respect to the time process. Switching integration variables according to $t-t_n=y$ and using $t_{n+1}-t_n=\tau_n$ we find
\begin{eqnarray}
C_n(s,u)&=&\eta^2c^2\left\langle e^{-t_n s} \tau_n^{2(\nu-\eta)}\int_0^{{\tau}_n} d y e^{-s y} y^{\eta-1}\int_0^{{\tau}_n-y} d \Delta e^{-u \Delta}(y+\Delta)^{\eta-1}\right\rangle    
\end{eqnarray}
From renewal assumption the random variables $t_n$ and $\tau_{n+1}$ are independent. Further since $t_n=\sum_{j=0}^{n-1}\tau_j$ and because the waiting times are also independent we have $\left\langle\exp \left(-s t_n\right)\right\rangle=\psi^n(s)$, where $\psi(s)$ is the Laplace $\tau \rightarrow s$ transform of $\psi(\tau)$. The remaining average is with respect to $\tau_n$, which is a random variable drawn from $\psi(\tau)$. Hence, we get
\begin{eqnarray}
C_n(s, u)&=&\eta^2c^2 \psi^n(s) \int_0^{\infty} d \tau \psi(\tau)\tau^{2(\nu-\eta)} \int_0^{\tau} d y e^{-s y} y^{\eta-1}\int_0^{\tau-y} d \Delta e^{-u \Delta}(y+\Delta)^{\eta-1}    
\end{eqnarray}

Let us denote
\begin{eqnarray}
W(\tau)=\int_{\tau}^{\infty} d \tilde{\tau} \psi\left(\tilde{\tau}\right)\tilde{\tau}^{2(\nu-\eta)}    
\end{eqnarray}

The above integral is convergent only if $2\nu\leq\gamma+2\eta$. Keeping that condition in mind, we can write

\begin{eqnarray}
C_n(s, u)=\eta^2c^2 \psi^n(s) \int_0^{\infty} d \tau \left[-\frac{d}{d \tau} W(\tau)\right]\int_0^{\tau} d y e^{-s y} y^{\eta-1} \int_0^{\tau-y} d \Delta e^{-u \Delta}(y+\Delta)^{\eta-1}    
\end{eqnarray}

We now integrate by parts, and then the geometric series $\sum_{n=0}^{\infty} \hat{\psi}^n(s)=1 /[1-\hat{\psi}(s)]$ gives
\begin{eqnarray}
C(s, u)=\frac{\eta^2c^2}{1-\psi(s)} \int_0^{\infty} d \tau W(\tau)\tau^{\eta-1} e^{-u \tau} \int_0^{\tau} d y e^{-(s-u) y} y^{\eta-1}
\end{eqnarray}
The double laplace transform of the above function will give us the velocity correlation function, $C(t,\Delta)$. We can invert $u$ to $\Delta$ in the above function, resulting
\begin{equation}
C(s,\Delta)=\frac{\eta^2c^2}{1-\psi(s)} \int_\Delta^{\infty} d \tau W(\tau)\tau^{\eta-1}(\tau-\Delta)^{\eta-1} e^{-s(\tau-\Delta)}    \end{equation}
Putting the value of $W(\tau)$,
\begin{equation}
C(s,\Delta)=\frac{\gamma t_0^\gamma\eta^2c^2}{(2(\nu-\eta)-\gamma)(1-\psi(s))}\int_\Delta^{\infty} d \tau \tau^{2\nu-\eta-\gamma-1}(\tau-\Delta)^{\eta-1} e^{-s(\tau-\Delta)}   \end{equation}

When $\eta=1$, the above equation would give us the expression of Generalized L\'evy walk 
\begin{eqnarray}
C(s, u)=\frac{\gamma t_0^\gamma\eta^2c^2}{(2(\nu-\eta)-\gamma)(1-\psi(s))} \frac{W(s)-W(u)}{u-s}    
\end{eqnarray}
where $W(s)$ is the Laplace transform of $W(\tau)$.
\subsection{$0<\gamma<1$}
In this region
\begin{equation}
    \psi(s)\simeq1-\Gamma(1-\gamma) t_0^\gamma s^\gamma
\end{equation}
$C(s,\Delta)$ would be expressed as 
\begin{eqnarray}
 C(s,\Delta)=\frac{\gamma\eta^2c^2}{(2(\nu-\eta)-\gamma)\Gamma(1-\gamma)s^\gamma}\int_\Delta^{\infty} d \tau \tau^{2\nu-\eta-\gamma-1}(\tau-\Delta)^{\eta-1} e^{-s(\tau-\Delta)}    
\end{eqnarray}
\subsubsection{$0<\gamma<1\hspace{.2cm}\&\hspace{.2cm}\gamma+1<2\nu<\gamma+2$}
For inverting $s\rightarrow t$, the above equation can be approximated for this region 
\begin{eqnarray}
 C(t,\Delta)=\frac{\gamma\eta^2c^2}{(2(\nu-\eta)-\gamma)\Gamma(1-\gamma)\Gamma(\gamma)}\int_\Delta^{t+\Delta} d \tau \tau^{2\nu-\eta-\gamma-1}(\tau-\Delta)^{\eta-1} [t-\tau-\Delta)]^{\gamma-1}   
\end{eqnarray}
Putting $p=\frac{\tau}{t}$ and $z=\frac{\Delta}{t}$
\begin{eqnarray}
 C(t,\Delta)=\frac{\gamma\eta^2c^2}{(2(\nu-\eta)-\gamma)\Gamma(1-\gamma)\Gamma(\gamma)}t^{2\nu-2}\int_z^{1+z} d p p^{2\nu-\eta-\gamma-1}(p-z)^{\eta-1}[1-(p-z)]^{\gamma-1}
\end{eqnarray}
The above integration is a scaling function with respect to $z$
\begin{eqnarray}
\phi(z)=\int_z^{1+z} d p p^{2\nu-\eta-\gamma-1}(p-z)^{\eta-1}[1-(p-z)]^{\gamma-1}    
\end{eqnarray}
Introducing $l=(p-z)^{-1}$ and $l^{\prime}=l^{-1}$, and considering $1+z l=\frac{t+\tau}{t} \rightarrow 1$ for large $t$, it can be expressed as
\begin{eqnarray}
\phi(z) &=&\int_1^{\infty} d l l^{1-\eta}(l-1)^{\eta-1}(1+z l)^{2\nu-\eta-\gamma-1}\nonumber \\
&=&B(2 \nu-\gamma-1, \eta)    
\end{eqnarray}

where $B(P, Q)=\int_0^1 d x x^{P-1}(1-x)^{Q-1}=\frac{\Gamma(P) \Gamma(Q)}{\Gamma(P+Q)}$ is the $\beta$ function.
Putting this in the expression of $C(t,\Delta)$ would give us
\begin{eqnarray}
  C(t, \Delta)=\frac{\gamma\eta^2c^2}{(2(\nu-\eta)-\gamma)\Gamma(1-\gamma)\Gamma(\gamma)} B(2 \nu-\gamma-1, \eta) t^{2 \nu-2}  
\end{eqnarray}

\subsubsection{$0<\gamma<1\hspace{.2cm}\&\hspace{.2cm}\gamma<2\nu<\gamma+1$}
In this region the behavior of the VCF in the long time limit $t \rightarrow \infty$, i.e., $s \rightarrow 0$ is obtained by utilizing the small $s$ expansion,
\begin{eqnarray}
e^{-s(\Delta-\tau)}=1-s(\Delta-\tau)+O\left(s^2\right) \simeq 1, s \rightarrow 0    
\end{eqnarray}
$C(s,\Delta)$ would be expressed as 
\begin{eqnarray}
 C(s,\Delta)=\frac{\gamma\eta^2c^2}{(2(\nu-\eta)-\gamma)\Gamma(1-\gamma)s^\gamma}\int_\Delta^{\infty} d \tau \tau^{2\nu-\eta-\gamma-1}(\tau-\Delta)^{\eta-1}   
\end{eqnarray}
Inversing $s$ to $t$ would change the above equation as 
\begin{eqnarray}
 C(t,\Delta)=\frac{\gamma\eta^2c^2t^{\gamma-1}}{(2(\nu-\eta)-\gamma)\Gamma(1-\gamma)\Gamma(\gamma)}\int_\Delta^{\infty} d \tau \tau^{2\nu-\eta-\gamma-1}(\tau-\Delta)^{\eta-1}   
\end{eqnarray}

Introducing $z=\frac{\Delta}{t}$,$r=\frac{\tau}{\Delta}$and$r^{\prime}=r^{-1}$, then above  equation can be expressed as
\begin{eqnarray}
 C(t,\Delta)=\frac{\gamma\eta^2c^2t^{2\gamma-2}z^{2\nu-\gamma-1}}{(2(\nu-\eta)-\gamma)\Gamma(1-\gamma)\Gamma(\gamma)}\int_1^{\infty} dr r^{2\nu-\eta-\gamma-1}(r-1)^{\eta-1}   
\end{eqnarray}

The integration in above equation is
\begin{eqnarray}
\phi(z) &=&\int_1^{\infty} d r r^{2\nu-\eta-\gamma-1}(r-1)^{\eta-1}\nonumber \\
&=&B(1+\gamma+\eta-2\nu, \eta)    
\end{eqnarray}

Then we have

\begin{eqnarray}
 C(t,\Delta)=\frac{\gamma\eta^2c^2}{(2(\nu-\eta)-\gamma)\Gamma(1-\gamma)\Gamma(\gamma)}B(1+\gamma+\eta-2\nu, \eta)t^{2\gamma-2}z^{2\nu-\gamma-1} 
\end{eqnarray}

\subsubsection{$0<\gamma<1\hspace{.2cm}\&\hspace{.2cm} 2\nu<\gamma$}
VCF for this region is same as the previous region. Therefore,
\begin{eqnarray}
 C(t,\Delta)=\frac{\gamma\eta^2c^2t^{2\gamma-2}z^{2\nu-\gamma-1}}{(2(\nu-\eta)-\gamma)\Gamma(1-\gamma)\Gamma(\gamma)}B(1+\gamma+\eta-2\nu, \eta) 
\end{eqnarray}
\subsection{$\gamma>1$}
In this region
\begin{equation}
    \psi(s)\simeq1-\langle\tau\rangle s+\Gamma(1-\gamma) t_0^\gamma s^\gamma
\end{equation}
Where $\langle\tau\rangle$ is $\frac{\gamma t_0}{\gamma-1}$.

$C(s,\Delta)$ would be expressed as 
\begin{eqnarray}
 C(s,\Delta)=\frac{\gamma\eta^2c^2}{(2(\nu-\eta)-\gamma)\langle\tau\rangle s}\int_\Delta^{\infty} d \tau \tau^{2\nu-\eta-\gamma-1}(\tau-\Delta)^{\eta-1} e^{-s(\tau-\Delta)}    
\end{eqnarray}
\subsubsection{$\gamma>1\hspace{.2cm}\&\hspace{.2cm}\gamma+1<2\nu<\gamma+2$}
For inverting $s\rightarrow t$, the above equation can be approximated for this region 
\begin{eqnarray}
 C(t,\Delta)=\frac{\gamma\eta^2c^2}{(2(\nu-\eta)-\gamma)\langle\tau\rangle}\int_\Delta^{t+\Delta} d \tau \tau^{2\nu-\eta-\gamma-1}(\tau-\Delta)^{\eta-1}  
\end{eqnarray}
Putting $z=\frac{\Delta}{t}$ and $r=\frac{\tau}{\Delta}$
\begin{eqnarray}
 C(t,\Delta)=\frac{\gamma\eta^2c^2}{(2(\nu-\eta)-\gamma)\langle\tau\rangle}t^{2\nu-\gamma-1}z^{2\nu-\gamma-1}\int_1^{1+\frac{1}{z}} dr r^{2\nu-\eta-\gamma-1}(r-1)^{\eta-1}
\end{eqnarray}
Putting $r^{\prime\prime}=r-1$, the above integration would be
\begin{eqnarray}
\phi(z)&=&z^{2\nu-\gamma-1}\int_1^{1+\frac{1}{z}} dr r^{2\nu-\eta-\gamma-1}(r-1)^{\eta-1}\nonumber\\
&=&\frac{1}{2\nu-\gamma-1}
\end{eqnarray}

This approximation is valid when both $t$ and $\tau$ being long but finite , which yields $r^{\prime \prime-1}=\frac{\Delta}{\tau-\Delta} \rightarrow 0$ for large $\tau$. The above integration converges only when $2 \gamma>$
$1+\beta$. 
Putting this in the expression of $C(t,\Delta)$ would give us
\begin{eqnarray}
  C(t, \Delta)=\frac{\gamma\eta^2c^2}{(2(\nu-\eta)-\gamma)\langle\tau\rangle}\frac{1}{2\nu-\gamma-1}t^{2\nu-\gamma-1}
\end{eqnarray}

\subsubsection{$\gamma>1\hspace{.2cm}\&\hspace{.2cm}\gamma<2\nu<\gamma+1$}
In this region the behavior of the VCF in the long time limit $t \rightarrow \infty$, i.e., $s \rightarrow 0$ is obtained by utilizing the small $s$ expansion,
\begin{eqnarray}
e^{-s(\Delta-\tau)}=1-s(\Delta-\tau)+O\left(s^2\right) \simeq 1, s \rightarrow 0    
\end{eqnarray}
$C(s,\Delta)$ would be expressed as 
\begin{eqnarray}
 C(s,\Delta)=\frac{\gamma\eta^2c^2}{(2(\nu-\eta)-\gamma)\langle\tau\rangle s}\int_\Delta^{\infty} d \tau \tau^{2\nu-\eta-\gamma-1}(\tau-\Delta)^{\eta-1} e^{-s(\tau-\Delta)}   
\end{eqnarray}
Inversing $s$ to $t$ would change the above equation as 
\begin{eqnarray}
C(t,\Delta)=\frac{\gamma\eta^2c^2}{(2(\nu-\eta)-\gamma)\langle\tau\rangle }\int_\Delta^{\infty} d \tau \tau^{2\nu-\eta-\gamma-1}(\tau-\Delta)^{\eta-1}   
\end{eqnarray}

Introducing $z=\frac{\Delta}{t}$, $r=\frac{\tau}{\Delta}$ and $r^{\prime}=r^{-1}$, then above  equation can be expressed as
\begin{eqnarray}
 C(t,\Delta)&=&\frac{\gamma\eta^2c^2}{(2(\nu-\eta)-\gamma)\langle\tau\rangle }t^{2\nu-\gamma-1}z^{2\nu-\gamma-1}\int_1^{\infty} d r r^{2\nu-\eta-\gamma-1}(r-1)^{\eta-1}\nonumber\\
 &=&\frac{\gamma\eta^2c^2}{(2(\nu-\eta)-\gamma)\langle\tau\rangle }B(1+\gamma-2\nu,\eta)t^{2\nu-\gamma-1}z^{2\nu-\gamma-1}
 \end{eqnarray}

\subsubsection{$\gamma>1\hspace{.2cm}\&\hspace{.2cm} 2\nu<\gamma$}
VCF for this region is same as the previous region. Therefore,
\begin{eqnarray}
 C(t,\Delta)=\frac{\gamma\eta^2c^2}{(2(\nu-\eta)-\gamma)\langle\tau\rangle }B(1+\gamma-2\nu,\eta)t^{2\nu-\gamma-1}z^{2\nu-\gamma-1}
\end{eqnarray}
\end{widetext}
\bibliography{levy_walk}

@Article{abry1998wavelet,
  author    = {Abry, Patrice and Veitch, Darryl},
  journal   = {IEEE Transactions on Information Theory},
  title     = {Wavelet analysis of long-range-dependent traffic},
  year      = {1998},
  issn      = {0018-9448},
  number    = {1},
  pages     = {2-15},
  volume    = {44},
  doi       = {10.1109/18.650984},
  publisher = {Institute of Electrical and Electronics Engineers (IEEE)},
}

@article{aghion2021moses,
  title={Moses, Noah and Joseph effects in L{\'e}vy walks},
  author={Aghion, Erez and Meyer, Philipp G and Adlakha, Vidushi and Kantz, Holger and Bassler, Kevin E},
  journal={New Journal of Physics},
  volume={23},
  number={2},
  pages={023002},
  year={2021},
  publisher={IOP Publishing}
}

@article{akimoto2020infinite,
  title={Infinite invariant density in a semi-Markov process with continuous state variables},
  author={Akimoto, Takuma and Barkai, Eli and Radons, G{\"u}nter},
  journal={Physical Review E},
  volume={101},
  number={5},
  pages={052112},
  year={2020},
  publisher={APS}
}

@Article{akimoto2013distributional,
  author    = {Akimoto, Takuma and Miyaguchi, Tomoshige},
  journal   = {Physical Review E—Statistical, Nonlinear, and Soft Matter Physics},
  title     = {Distributional ergodicity in stored-energy-driven l{\'e}vy flights},
  year      = {2013},
  number    = {6},
  pages     = {062134},
  volume    = {87},
  publisher = {APS},
}

@Article{akimoto2014phase,
  author    = {Akimoto, Takuma and Miyaguchi, Tomoshige},
  journal   = {Journal of Statistical Physics},
  title     = {Phase diagram in stored-energy-driven L{\'e}vy flight},
  year      = {2014},
  pages     = {515--530},
  volume    = {157},
  publisher = {Springer},
}

@Article{albers2018exact,
  author    = {Albers, Tony and Radons, G{\"u}nter},
  journal   = {Physical Review Letters},
  title     = {Exact Results for the Nonergodicity of d-Dimensional Generalized Lévy Walks},
  year      = {2018},
  issn      = {0031-9007},
  number    = {10},
  pages     = {104501},
  volume    = {120},
  doi       = {10.1103/physrevlett.120.104501},
  publisher = {American Physical Society (APS)},
}

@Article{albers2022nonergodicity,
  author    = {Albers, Tony and Radons, G{\"u}nter},
  journal   = {Physical Review E},
  title     = {Nonergodicity of d-dimensional generalized L{\'e}vy walks and their relation to other space-time coupled models},
  year      = {2022},
  number    = {1},
  pages     = {014113},
  volume    = {105},
  publisher = {APS},
}

@Book{benkadda1998chaos,
  author    = {Benkadda, Sadruddin and Zaslavsky, George M},
  publisher = {Springer Science \& Business Media},
  title     = {Chaos, Kinetics and Nonlinear Dynamics in Fluids and Plasmas: Proceedings of a Workshop Held in Carry-Le Rouet, France, 16--21 June 1997},
  year      = {1998},
  volume    = {511},
}

@article{bassler2007nonstationary,
  title={Nonstationary increments, scaling distributions, and variable diffusion processes in financial markets},
  author={Bassler, Kevin E and McCauley, Joseph L and Gunaratne, Gemunu H},
  journal={Proceedings of the National Academy of Sciences},
  volume={104},
  number={44},
  pages={17287--17290},
  year={2007},
  publisher={National Academy of Sciences}
}

@Article{bothe2019mean,
  author    = {Bothe, Marius and Sagues, Francesc and Sokolov, Igor M},
  journal   = {Physical Review E},
  title     = {Mean squared displacement in a generalized L{\'e}vy walk model},
  year      = {2019},
  number    = {1},
  pages     = {012117},
  volume    = {100},
  publisher = {APS},
}

@Article{godreche2001statistics,
  author    = {Godreche, C and Luck, JM1853425},
  journal   = {Journal of Statistical Physics},
  title     = {Statistics of the occupation time of renewal processes},
  year      = {2001},
  pages     = {489--524},
  volume    = {104},
  publisher = {Springer},
}

@Book{mandelbrot2002gaussian,
  author    = {Mandelbrot, Benoit},
  publisher = {Springer Science \& Business Media},
  title     = {Gaussian self-affinity and fractals: globality, the earth, 1/f noise, and R/S},
  year      = {2002},
}

@Article{zaburdaev2015levy,
  author    = {Zaburdaev, Vasily and Denisov, Sergey and Klafter, Joseph},
  journal   = {Reviews of Modern Physics},
  title     = {L{\'e}vy walks},
  year      = {2015},
  number    = {2},
  pages     = {483--530},
  volume    = {87},
  publisher = {APS},
}

@Article{oliveira2019anomalous,
  author    = {Oliveira, Fernando A and Ferreira, Rogelma MS and Lapas, Luciano C and Vainstein, Mendeli H},
  journal   = {Frontiers in Physics},
  title     = {Anomalous diffusion: A basic mechanism for the evolution of inhomogeneous systems},
  year      = {2019},
  pages     = {18},
  volume    = {7},
  publisher = {Frontiers Media SA},
}

@Article{metzler2014anomalous,
  author    = {Metzler, Ralf and Jeon, Jae-Hyung and Cherstvy, Andrey G and Barkai, Eli},
  journal   = {Physical Chemistry Chemical Physics},
  title     = {Anomalous diffusion models and their properties: non-stationarity, non-ergodicity, and ageing at the centenary of single particle tracking},
  year      = {2014},
  number    = {44},
  pages     = {24128--24164},
  volume    = {16},
  publisher = {Royal Society of Chemistry},
}

@Article{hofling2013anomalous,
  author    = {H{\"o}fling, Felix and Franosch, Thomas},
  journal   = {Reports on Progress in Physics},
  title     = {Anomalous transport in the crowded world of biological cells},
  year      = {2013},
  number    = {4},
  pages     = {046602},
  volume    = {76},
  publisher = {IOP Publishing},
}

@Article{chen2017anomalous,
  author    = {Chen, Lijian and Bassler, Kevin E and McCauley, Joseph L and Gunaratne, Gemunu H},
  journal   = {Physical Review E},
  title     = {Anomalous scaling of stochastic processes and the Moses effect},
  year      = {2017},
  number    = {4},
  pages     = {042141},
  volume    = {95},
  publisher = {APS},
}

@Article{meyer2018anomalous,
  author    = {Meyer, Philipp G and Adlakha, Vidushi and Kantz, Holger and Bassler, Kevin E},
  journal   = {New Journal of Physics},
  title     = {Anomalous diffusion and the Moses effect in an aging deterministic model},
  year      = {2018},
  number    = {11},
  pages     = {113033},
  volume    = {20},
  publisher = {IOP Publishing},
}

@Article{meyer2022decomposing,
  author    = {Meyer, Philipp G and Aghion, Erez and Kantz, Holger},
  journal   = {Journal of Physics A: Mathematical and Theoretical},
  title     = {Decomposing the effect of anomalous diffusion enables direct calculation of the Hurst exponent and model classification for single random paths},
  year      = {2022},
  number    = {27},
  pages     = {274001},
  volume    = {55},
  publisher = {IOP Publishing},
}

@Article{hurst1951long,
  author    = {Hurst, Harold Edwin},
  journal   = {Transactions of the American Society of Civil Engineers},
  title     = {Long-term storage capacity of reservoirs},
  year      = {1951},
  number    = {1},
  pages     = {770--799},
  volume    = {116},
  publisher = {American Society of Civil Engineers},
}

@Article{alexander1969comments,
  author    = {Alexander, GN},
  journal   = {Water Resources Research},
  title     = {Comments on ‘Noah, Joseph, and Operational Hydrology’ by Benoit B. Mandelbrot and James R. Wallis},
  year      = {1969},
  issn      = {0043-1397},
  number    = {4},
  pages     = {915-916},
  volume    = {5},
  doi       = {10.1029/wr005i004p00915},
  publisher = {American Geophysical Union (AGU)},
}

@Article{peng1994mosaic,
  author    = {Peng, C-K and Buldyrev, Sergey V and Havlin, Shlomo and Simons, Michael and Stanley, H Eugene and Goldberger, Ary L},
  journal   = {Physical Review E},
  title     = {Mosaic organization of DNA nucleotides},
  year      = {1994},
  number    = {2},
  pages     = {1685},
  volume    = {49},
  publisher = {APS},
}

@article{vilk2022unravelling,
  title={Unravelling the origins of anomalous diffusion: from molecules to migrating storks},
  author={Vilk, Ohad and Aghion, Erez and Avgar, Tal and Beta, Carsten and Nagel, Oliver and Sabri, Adal and Sarfati, Raphael and Schwartz, Daniel K and Weiss, Matthias and Krapf, Diego and others},
  journal={Physical Review Research},
  volume={4},
  number={3},
  pages={033055},
  year={2022},
  publisher={APS}
}

@article{montroll1965random,
  title={Random walks on lattices. II},
  author={Montroll, Elliott W and Weiss, George H},
  journal={Journal of Mathematical Physics},
  volume={6},
  number={2},
  pages={167--181},
  year={1965},
  publisher={American Institute of Physics}
}

@article{vezzani2020rare,
  title={Rare events in generalized L{\'e}vy walks and the big jump principle},
  author={Vezzani, Alessandro and Barkai, Eli and Burioni, Raffaella},
  journal={Scientific Reports},
  volume={10},
  number={1},
  pages={2732},
  year={2020},
  publisher={Nature Publishing Group UK London}
}

@article{metzler2019brownian,
  title={Brownian motion and beyond: first-passage, power spectrum, non-Gaussianity, and anomalous diffusion},
  author={Metzler, Ralf},
  journal={Journal of Statistical Mechanics: Theory and Experiment},
  volume={2019},
  number={11},
  pages={114003},
  year={2019},
  publisher={IOP Publishing}
}

@article{sabri2020elucidating,
  title={Elucidating the origin of heterogeneous anomalous diffusion in the cytoplasm of mammalian cells},
  author={Sabri, Adal and Xu, Xinran and Krapf, Diego and Weiss, Matthias},
  journal={Physical Review Letters},
  volume={125},
  number={5},
  pages={058101},
  year={2020},
  publisher={APS}
}

@Article{seemann2012ensemble,
  author    = {Seemann, Lars and Hua, Jia-Chen and McCauley, Joseph L and Gunaratne, Gemunu H},
  journal   = {Physica A: Statistical Mechanics and its Applications},
  title     = {Ensemble vs. time averages in financial time series analysis},
  year      = {2012},
  issn      = {0378-4371},
  number    = {23},
  pages     = {6024-6032},
  volume    = {391},
  doi       = {10.1016/j.physa.2012.06.054},
  publisher = {Elsevier BV},
}

@article{lim2002self,
  title={Self-similar Gaussian processes for modeling anomalous diffusion},
  author={Lim, Soonchieh C and Muniandy, Sithi Vinayakam},
  journal={Physical Review E},
  volume={66},
  number={2},
  pages={021114},
  year={2002},
  publisher={APS}
}

@article{jeon2014scaled,
  title={Scaled Brownian motion: a paradoxical process with a time dependent diffusivity for the description of anomalous diffusion},
  author={Jeon, Jae-Hyung and Chechkin, Aleksei V and Metzler, Ralf},
  journal={Physical Chemistry Chemical Physics},
  volume={16},
  number={30},
  pages={15811--15817},
  year={2014},
  publisher={Royal Society of Chemistry}
}

@article{thiel2014scaled,
  title={Scaled Brownian motion as a mean-field model for continuous-time random walks},
  author={Thiel, Felix and Sokolov, Igor M},
  journal={Physical Review E},
  volume={89},
  number={1},
  pages={012115},
  year={2014},
  publisher={APS}
}

@article{shlesinger1987levy,
  title={L{\'e}vy dynamics of enhanced diffusion: Application to turbulence},
  author={Shlesinger, Michael F and West, BJ and Klafter, Joseph},
  journal={Physical Review Letters},
  volume={58},
  number={11},
  pages={1100},
  year={1987},
  publisher={APS}
}

@article{meyer2017scale,
  title={Scale-invariant Green-Kubo relation for time-averaged diffusivity},
  author={Meyer, Philipp and Barkai, Eli and Kantz, Holger},
  journal={Physical Review E},
  volume={96},
  number={6},
  pages={062122},
  year={2017},
  publisher={APS}
}

@article{PhysRevX.4.011022,
  title = {Scaling Green-Kubo Relation and Application to Three Aging Systems},
  author = {Dechant, A. and Lutz, E. and Kessler, D. A. and Barkai, E.},
  journal = {Phys. Rev. X},
  volume = {4},
  issue = {1},
  pages = {011022},
  numpages = {16},
  year = {2014},
  month = {Feb},
  publisher = {American Physical Society},
  doi = {10.1103/PhysRevX.4.011022},
  url = {https://link.aps.org/doi/10.1103/PhysRevX.4.011022}
}

@Article{schulz1997anomalous,
  author    = {Schulz-Baldes, Hermann},
  journal   = {Physical Review Letters},
  title     = {Anomalous Drude Model},
  year      = {1997},
  issn      = {0031-9007},
  number    = {11},
  pages     = {2176-2179},
  volume    = {78},
  doi       = {10.1103/physrevlett.78.2176},
  publisher = {American Physical Society (APS)},
}

@Article{dechant2012anomalous,
  author    = {Dechant, Andreas and Lutz, Eric},
  journal   = {Physical Review Letters},
  title     = {Anomalous Spatial Diffusion and Multifractality in Optical Lattices},
  year      = {2012},
  issn      = {0031-9007},
  number    = {23},
  pages     = {230601},
  volume    = {108},
  doi       = {10.1103/physrevlett.108.230601},
  publisher = {American Physical Society (APS)},
}

@Article{plakhotnik2010anomalous,
  author    = {Plakhotnik, Taras and Fern{\'e}e, Mark J and Littleton, Brad and Rubinsztein-Dunlop, Halina and Potzner, Christian and Mulvaney, Paul},
  journal   = {Physical Review Letters},
  title     = {Anomalous Power Laws of Spectral Diffusion in Quantum Dots: A Connection to Luminescence Intermittency},
  year      = {2010},
  issn      = {0031-9007},
  number    = {16},
  pages     = {167402},
  volume    = {105},
  doi       = {10.1103/physrevlett.105.167402},
  publisher = {American Physical Society (APS)},
}

@Article{margolin2004aging,
  author    = {Margolin, Gennady and Barkai, Eli},
  journal   = {The Journal of Chemical Physics},
  title     = {Aging correlation functions for blinking nanocrystals, and other on–off stochastic processes},
  year      = {2004},
  issn      = {0021-9606},
  number    = {3},
  pages     = {1566--1577},
  volume    = {121},
  doi       = {10.1063/1.1763136},
  publisher = {AIP Publishing},
}

@Article{laplace1810approximations,
  author  = {Laplace, Pierre-Simon},
  journal = {{\OE}uvres Compl{\`e}tes},
  title   = {Sur les approximations des formules qui sont fonctions de tres grands nombres et sur leur application aux probabilites},
  year    = {1810},
  pages   = {301--345},
  volume  = {12},
}

@book{brown2022miscellaneous,
  title={The Miscellaneous Botanical Works of Robert Brown: Vol. I},
  author={Brown, Robert},
  year={2022},
  publisher={BoD--Books on Demand}
}

@article{shlesinger1982random,
  title={Random walks with infinite spatial and temporal moments},
  author={Shlesinger, Michael F and Klafter, Joseph and Wong, YM},
  journal={Journal of Statistical Physics},
  volume={27},
  pages={499--512},
  year={1982},
  publisher={Springer}
}

@Article{klafter1996beyond,
  author    = {Klafter, Joseph and Shlesinger, Michael F and Zumofen, Gert},
  journal   = {Physics Today},
  title     = {Beyond brownian motion},
  year      = {1996},
  number    = {2},
  pages     = {33--39},
  volume    = {49},
  publisher = {American Institute of Physics},
}

@article{shlesinger1993strange,
  title={Strange kinetics},
  author={Shlesinger, Michael F and Zaslavsky, George M and Klafter, Joseph},
  journal={Nature},
  volume={363},
  number={6424},
  pages={31--37},
  year={1993},
  publisher={Nature Publishing Group UK London}
}

@article{zumofen1993levy,
  title={L{\'e}vy walks and propagators in intermittent chaotic systems},
  author={Zumofen, G and Klafter, J and Blumen, A},
  journal={Physical Review E},
  volume={47},
  number={3},
  pages={2183},
  year={1993},
  publisher={APS}
}

@article{zumofen1993scale,
  title={Scale-invariant motion in intermittent chaotic systems},
  author={Zumofen, G and Klafter, J},
  journal={Physical Review E},
  volume={47},
  number={2},
  pages={851},
  year={1993},
  publisher={APS}
}

@article{klafter1987stochastic,
  title={Stochastic pathway to anomalous diffusion},
  author={Klafter, Joseph and Blumen, Alexander and Shlesinger, Michael F},
  journal={Physical Review A},
  volume={35},
  number={7},
  pages={3081},
  year={1987},
  publisher={APS}
}

@Misc{davey1991mandelbrot,
  author    = {Benoit B. Mandelbrot},
  title     = {Fractal Geometry in Physics},
  year      = {1993},
  doi       = {10.21236/ada273271},
  journal   = {Earth SciencesDivision},
  pages     = {40},
  publisher = {Defense Technical Information Center},
}

@article{hu2025generalized,
  title={Generalized two-state random walk model: Nontrivial anomalous diffusion, aging, and ergodicity breaking},
  author={Hu, Yuhang and Liu, Jian},
  journal={Physical Review E},
  volume={111},
  number={1},
  pages={014148},
  year={2025},
  publisher={APS}
}

@article{del2000chaotic,
  title={Chaotic transport in zonal flows in analogous geophysical and plasma systems},
  author={del-Castillo-Negrete, Diego},
  journal={Physics of Plasmas},
  volume={7},
  number={5},
  pages={1702--1711},
  year={2000},
  publisher={American Institute of Physics}
}

@article{del1998asymmetric,
  title={Asymmetric transport and non-Gaussian statistics of passive scalars in vortices in shear},
  author={del-Castillo-Negrete, D},
  journal={Physics of Fluids},
  volume={10},
  number={3},
  pages={576--594},
  year={1998},
  publisher={American Institute of Physics}
}

@article{zaburdaev2011perturbation,
  title={Perturbation spreading in many-particle systems: a random walk approach},
  author={Zaburdaev, Vasily and Denisov, S and H{\"a}nggi, Peter},
  journal={Physical review letters},
  volume={106},
  number={18},
  pages={180601},
  year={2011},
  publisher={APS}
}

@article{sims2008scaling,
  title={Scaling laws of marine predator search behaviour},
  author={Sims, David W and Southall, Emily J and Humphries, Nicolas E and Hays, Graeme C and Bradshaw, Corey JA and Pitchford, Jonathan W and James, Alex and Ahmed, Mohammed Z and Brierley, Andrew S and Hindell, Mark A and others},
  journal={Nature},
  volume={451},
  number={7182},
  pages={1098--1102},
  year={2008},
  publisher={Nature Publishing Group UK London}
}

@article{berkowitz2016measurements,
  title={Measurements and models of reactive transport in geological media},
  author={Berkowitz, Brian and Dror, Ishai and Hansen, Scott K and Scher, Harvey},
  journal={Reviews of Geophysics},
  volume={54},
  number={4},
  pages={930--986},
  year={2016},
  publisher={Wiley Online Library}
}

@article{cortis2004anomalous,
  title={Anomalous transport in “classical” soil and sand columns},
  author={Cortis, Andrea and Berkowitz, Brian},
  journal={Soil Science Society of America Journal},
  volume={68},
  number={5},
  pages={1539--1548},
  year={2004},
  publisher={Wiley Online Library}
}

@article{levy2003measurement,
  title={Measurement and analysis of non-Fickian dispersion in heterogeneous porous media},
  author={Levy, Melissa and Berkowitz, Brian},
  journal={Journal of contaminant hydrology},
  volume={64},
  number={3-4},
  pages={203--226},
  year={2003},
  publisher={Elsevier}
}

@article{bijeljic2011signature,
  title={Signature of non-Fickian solute transport in complex heterogeneous porous media},
  author={Bijeljic, Branko and Mostaghimi, Peyman and Blunt, Martin J},
  journal={Physical review letters},
  volume={107},
  number={20},
  pages={204502},
  year={2011},
  publisher={APS}
}

@misc{supp,
  note = "See Supplemental Material at http://link.aps.org/
supplemental/10.1103/PhysRevLett.120.104501."
}

@article{munoz2021objective,
  title={Objective comparison of methods to decode anomalous diffusion},
  author={Mu{\~n}oz-Gil, Gorka and Volpe, Giovanni and Garcia-March, Miguel Angel and Aghion, Erez and Argun, Aykut and Hong, Chang Beom and Bland, Tom and Bo, Stefano and Conejero, J Alberto and Firbas, Nicol{\'a}s and others},
  journal={Nature communications},
  volume={12},
  number={1},
  pages={6253},
  year={2021},
  publisher={Nature Publishing Group UK London}
}

@article{argun2021classification,
  title={Classification, inference and segmentation of anomalous diffusion with recurrent neural networks},
  author={Argun, Aykut and Volpe, Giovanni and Bo, Stefano},
  journal={Journal of Physics A: Mathematical and Theoretical},
  volume={54},
  number={29},
  pages={294003},
  year={2021},
  publisher={IOP Publishing}
}

@article{garibo2021efficient,
  title={Efficient recurrent neural network methods for anomalously diffusing single particle short and noisy trajectories},
  author={Garibo-i-Orts, {\`O}scar and Baeza-Bosca, Alba and Garcia-March, Miguel A and Conejero, J Alberto},
  journal={Journal of Physics A: Mathematical and Theoretical},
  volume={54},
  number={50},
  pages={504002},
  year={2021},
  publisher={IOP Publishing}
}

@article{malinowski2025cinnamon,
  title={CINNAMON: A hybrid approach to change point detection and parameter estimation in single-particle tracking data},
  author={Malinowski, Jakub and Kostrzewa, Marcin and Balcerek, Micha{\l} and Tomczuk, Weronika and Szwabi{\'n}ski, Janusz},
  journal={Journal of Physics: Photonics},
  volume={7},
  number={3},
  pages={035008},
  year={2025},
  publisher={IOP Publishing}
}

@article{trillot2025evidence,
  title={Evidence and origin of anomalous diffusion of Ag+ ion in amorphous silica: a molecular dynamics study with neural network interatomic potentials},
  author={Trillot, Salom{\'e} and Tarrat, Nathalie and Combe, Nicolas and Benzo, Patrizio and Bonafos, Caroline and Benoit, Magali},
  journal={The Journal of Chemical Physics},
  volume={162},
  number={10},
  year={2025},
  publisher={AIP Publishing}
}

@article{meyer2023return,
  title={Return over volume statistics and the Moses effect in S\&P 500 data},
  author={Meyer, Philipp G and Zamani, Maryam and Kantz, Holger},
  journal={Physica A: Statistical Mechanics and its Applications},
  volume={612},
  pages={128497},
  year={2023},
  publisher={Elsevier}
}

@article{barraza2025non,
  title={A non-homogeneous, non-stationary and path-dependent Markov anomalous diffusion model},
  author={Barraza, Nestor R and Pena, Gabriel and Gambini, Juliana and Carusela, M Florencia},
  journal={Journal of Physics A: Mathematical and Theoretical},
  volume={58},
  number={9},
  pages={095001},
  year={2025},
  publisher={IOP Publishing}
}

@article{zamani2021anomalous,
  title={Anomalous diffusion in the citation time series of scientific publications},
  author={Zamani, Maryam and Aghion, Erez and Pollner, Peter and Vicsek, Tamas and Kantz, Holger},
  journal={Journal of Physics: Complexity},
  volume={2},
  number={3},
  pages={035024},
  year={2021},
  publisher={IOP Publishing}
}

@phdthesis{salek2024statistical,
  title={Statistical analysis and modeling of the opening and closing auctions of financial markets},
  author={Salek, Mohammed},
  year={2024},
  school={Universit{\'e} Paris-Saclay}
}

@book{fischer2011history,
  title={A history of the central limit theorem: from classical to modern probability theory},
  author={Fischer, Hans},
  volume={4},
  year={2011},
  publisher={Springer}
}

@article{salek2024equity,
  title={Equity auction dynamics: latent liquidity models with activity acceleration},
  author={Salek, Mohammed and Challet, Damien and Muni Toke, Ioane},
  journal={Quantitative Finance},
  volume={24},
  number={10},
  pages={1381--1398},
  year={2024},
  publisher={Taylor \& Francis}
}
\end{document}